\documentclass[times,twocolumn,final]{elsarticle}

\usepackage{medima}
\usepackage{framed,multirow}
\usepackage{pifont}
\usepackage{rotating}
\usepackage{color}
\usepackage{todonotes}

\usepackage{amssymb}
\usepackage{latexsym}
\usepackage{amsmath}
\usepackage{url}
\usepackage{xcolor}

\usepackage{hyperref}

\definecolor{newcolor}{rgb}{.8,.349,.1}

\journal{Preprint}

\begin{document}

\verso{Zhen Yuan \textit{et~al.}}

\begin{frontmatter}

\title{DeepSPV: A Deep Learning Pipeline for 3D Spleen Volume Estimation from 2D Ultrasound Images}%

\author[1]{Zhen \snm{Yuan}\corref{cor1}}
\cortext[cor1]{Corresponding author: 
  zhen.1.yuan@kcl.ac.uk}

\author[1]{David \snm{Stojanovski}}

\author[2,3]{Lei \snm{Li}}

\author[4,1]{Alberto \snm{Gomez}}

\author[5]{Haran \snm{Jogeesvaran}}

\author[1]{Esther \snm{Puyol-Antón}}

\author[5]{Baba \snm{Inusa}}

\author[1]{Andrew P. \snm{King}}

\address[1]{School of Biomedical Engineering and Imaging Science, King's College London, London, UK}
\address[2]{Department of Biomedical Engineering, National University of Singapore, Singapore}
\address[3]{School of Electronics \& Computer Science, University of Southampton, Southampton, UK}
\address[4]{Ultromics Ltd, Oxford, UK}
\address[5]{Evelina London Children's Hospital, Guy's and St Thomas NHS Foundation Trust, London, UK}



\begin{abstract}
Splenomegaly, the enlargement of the spleen, is an important clinical indicator for various associated medical conditions, such as sickle cell disease (SCD). Spleen length measured from 2D ultrasound is the most widely used metric for characterising spleen size. However, it is still considered a surrogate measure, and spleen volume remains the gold standard for assessing spleen size. Accurate spleen volume measurement typically requires 3D imaging modalities, such as computed tomography or magnetic resonance imaging, but these are not widely available, especially in the Global South which has a high prevalence of SCD. In this work, we introduce a deep learning pipeline, DeepSPV, for precise spleen volume estimation from single or dual 2D ultrasound images. The pipeline involves a segmentation network and a variational autoencoder for learning low-dimensional representations from the estimated segmentations. We investigate three approaches for spleen volume estimation and our best model achieves 86.62\%/92.5\% mean relative volume accuracy (MRVA) under single-view/dual-view settings, surpassing the performance of human experts. In addition, the pipeline can provide confidence intervals for the volume estimates as well as offering benefits in terms of interpretability, which further support clinicians in decision-making when identifying splenomegaly. We evaluate the full pipeline using a highly realistic synthetic dataset generated by a diffusion model, achieving an overall MRVA of 83.0\% from a single 2D ultrasound image. Our proposed DeepSPV is the first work to use deep learning to estimate 3D spleen volume from 2D ultrasound images and can be seamlessly integrated into the current clinical workflow for spleen assessment. We also make our synthetic spleen ultrasound dataset publicly available.
\end{abstract}


\end{frontmatter}


\section{Introduction}
\label{Section 1}
\subsection{Clinical Background}
\label{Section 1.1}
Splenomegaly refers to an enlargement of the spleen and is an important clinical indicator for various underlying conditions, such as sickle cell disease (SCD) \citep{inusa2016introductory}. In paediatric patients with SCD, splenomegaly can be accompanied by acute splenic sequestration, in which excessive blood cells become trapped in the spleen, causing a significant drop in circulating blood volume and haemoglobin \citep{brousse2014spleen}. This condition can be life-threatening without timely intervention. Other clinical conditions, such as cancer \citep{mccormick2000splenomegaly}, blood diseases \citep{pozo2009splenomegaly}, and infections \citep{mahmoud1972mechanisms} can also lead to splenomegaly. Therefore, for SCD and other conditions, the size of the spleen is commonly measured as supportive evidence for determining patient management. This creates a need for methods to accurately measure spleen size and identify splenomegaly.

\subsection{Challenges}
\label{Section 1.2}
In clinical settings, a common initial approach to detect splenomegaly is manual palpation. If the spleen is palpable below the last left rib, it is suspected to be enlarged. While the examination result from manual palpation was previously considered to be evidence of splenic enlargement \citep{schloesser1963diagnostic}, later research revealed that up to 16\% of palpable spleens were of normal size upon radiological assessment \citep{arkles1986palpable}. Besides, manual palpation is coarse, non-quantitative, and provides only a preliminary judgement as to whether further imaging-based examination is required.

Ultrasound (US) is the most used imaging modality for detecting splenomegaly, and is commonly arranged after manual palpation at a clinical check-up. It has wide availability in most clinical settings, and is non-invasive and portable. The length of the spleen can be measured from 2D coronal US images and is commonly used as a measure of the overall spleen size \citep{pozo2009splenomegaly}, as it correlates well with spleen volume \citep{lamb2002spleen}. However, despite this correlation, spleen length is still a surrogate measure, and spleen volume remains the gold standard measure for spleen size. In addition, apart from determining splenomegaly, spleen volume is the preferred indicator in determining the severity of related diseases and informing treatment planning \citep{kotlyar2014spleen,koga2016splenic,khoshpouri2018correlation}.

3D imaging modalities, such as computed tomography (CT) and magnetic resonance imaging (MRI), are considered to be the gold standard methods for spleen volume measurement \citep{yetter2003estimating,holmstrom2022test,paul2017splenic, holmstrom2022test}. Based on a CT or MRI examination, spleen volume can be computed by manually segmenting the spleen, but this approach is prohibitively time-consuming. More commonly, it can be estimated from manual measurements of spleen dimensions followed by use of the linear regression formula proposed in \cite{prassopoulos1997determination}. Deep learning-based frameworks have also been proposed to perform spleen segmentation from 3D CT and MRI \citep{ahn2020deep,moon2019acceleration,humpire2020fully, altini2022liver,meddeb2021evaluation,altini2022liver,huo2018splenomegaly}, which also facilitate spleen volume measurements. Nevertheless, CT involves ionising radiation so may not be performed for spleen size estimation in many parts of the world, and reliable MRI images are often challenging to acquire due to the problem of motion-induced artefacts. In addition, because of their expense and complexity of installation and maintenance, 3D imaging modalities are less accessible in areas where there is a high prevalence of SCD-related splenomegaly, i.e., the Global South \citep{piel2013global,grosse2011sickle}.

3D US imaging is a potential alternative that could provide volumetric measurements directly. However, despite 3D US having been successfully applied to superficial and static anatomical structures such as lower-limb muscles \citep{al2021ifss,huet2024validation}, acquiring high-quality 3D US images of the entire spleen remains challenging. This is primarily due to the obstruction by the ribs and interference from the diaphragm or gas in the stomach/bowel which compromise image quality. In addition, it is not always possible to image the whole spleen in a single view due to the limited acoustic windows. Similar challenges have been reported in 3D US imaging for large organs such as the liver \citep{treece20013d}. 3D US may also be feasible using either sweeping 2D US acquisitions or compounding of freehand 2D US images. Traditional reconstruction approaches include sensor-based methods that rely on external tracking systems \citep{rohling1999comparison, daoud2015freehand, wen2013accurate}, as well as sensorless ones such as \cite{gee2006sensorless}. Recent advances in deep learning have further facilitated freehand 2D US reconstruction by enabling end-to-end training and inference, as demonstrated in \cite{luo2023recon, yeung2024sensorless, gaits2024ultrasound}. However, acquiring multiple (more than two) 2D US images from different angles for subsequent 3D reconstruction remains challenging due to the the potential for respiration or bulk motion-induced artefacts as well as the same limiting factors affecting 3D US acquisition discussed above. These limitations make it impractical to collect a series of 2D US images with consistent quality for reliable 3D reconstruction.

Given these limitations, 2D US imaging remains the standard modality in current clinical workflows for spleen assessment \citep{lamb2002spleen}. While 3D US and 3D reconstruction from 2D US remain impractical, spleen volume can be estimated using linear regression formulas based on measurements from coronal and transverse 2D US, including the standard prolate ellipsoid formula \citep{de1999normal, chow2016spleen} or its modified version \citep{yetter2003estimating}. These methods have been applied in clinical practice; however, their accuracy and robustness heavily depend on the experience of the observer. Moreover, such expertise is often lacking in low- and middle-income countries where SCD is most prevalent. Thus, measurement of spleen length from 2D US images is still the most widely adopted approach for spleen size assessment, and the clinical value of spleen volume as a biomarker is still under-investigated.

\subsection{Motivations and Contributions}
\label{1.3}
Given the challenges associated with CT, MRI, and 3D US, our overall objective was to develop a pipeline that can estimate spleen volume directly from 2D US images acquired using standard clinical views, in alignment with the current clinical workflow for spleen size examination. By `standard clinical views', we refer to the coronal and transverse views typically acquired at US examination. 

We propose DeepSPV, a novel deep learning-based pipeline for precise spleen volume estimation from single-view (coronal) or dual-view (coronal and transverse) 2D US images. The pipeline consists of two parts: first, automatic segmentation of the spleen from the 2D US images, and second, estimating volume from the segmentations. We successfully evaluated the full pipeline under the single-view condition using a dataset of highly realistic synthetic US images, generated from CT images (with ground truth spleen volumes) by applying an US semantic diffusion model (USDM). To the best of our knowledge, this is the first work to use deep learning for spleen volume estimation from 2D US images. The main contributions of this work can be summarised as: 
\begin{enumerate}
    \item We propose a novel variational autoencoder (VAE) based framework that can automatically estimate 3D spleen volume from single or dual 2D spleen segmentation(s).
    \item We propose and evaluate three different volume estimation methods within the framework.  
    \item As well as making an estimate of volume, our framework also allows estimation of confidence intervals and provides a certain level of interpretability to support decision-making in real clinical scenarios. 
    \item We produce and make publicly available a database of highly realistic synthetic US images with ground truth spleen volumes.
    \item Leveraging this database, we successfully evaluate the full pipeline under the single-view condition, achieving volume estimation accuracy that surpasses the performance of human experts using the same data.
\end{enumerate}

\section{Related Works}
\label{Section 2}
\subsection{Spleen Segmentation from US}
Building on developments in deep learning, and specifically convolutional neural networks (CNNs), the U-Net architecture and its various adaptations have been widely employed to automate segmentation for medical images acquired from different modalities. For 2D US, models based upon the U-Net have demonstrated exceptional performance, for example in breast \citep{zhuang2019nipple} and fetal \citep{qiao2020dilated} US images. More recently, foundation models have been proposed to enable cross-modality and cross-organ medical image segmentation using a single generalisable model \citep{ma2024segment, huang2024segment}. These models have the potential to mitigate the need for manual annotations to train an application-specific segmentation model. However, currently they may not reach the same level of performance as a model trained using data specific to the required task. Furthermore, whilst their generalisability has been demonstrated across various datasets, their performance on spleen ultrasound images has not yet been investigated. To date, the only work to have demonstrated deep learning-based spleen segmentation from 2D US was our previous work \citep{yuan2020deep,yuan2022deep} built based on the U-Net.

\subsection{Deep Learning-based Volume Estimation}
Deep learning-based methods have been widely studied for automated volume measurement, through segmentation-based methods and regression-based methods. In segmentation-based methods, volume is typically calculated from automatic segmentation of 3D imaging such as CT and MRI. This technique has been applied in organs including the heart \citep{liao2017estimation}, liver \citep{wang2019automated}, and kidney \citep{sharma2017automatic}. Similar approaches have also been explored for 3D US segmentation in volume estimation, such as for the placenta \citep{looney2018fully}. Apart from segmentation, direct volume regression has been attempted for the heart \citep{zhen2015direct} and the kidney \citep{hussain2021cascaded}, again from 3D imaging. In the spleen, 3D CNNs have been proposed by \cite{moon2019acceleration,humpire2020fully} to segment the spleen from 3D CT images and subsequently estimate its volume, whilst \cite{huo2018splenomegaly} focused on spleen segmentation from MRI images of splenomegaly cases.

All the works mentioned above require 3D modalities for volume estimation. However, 3D CT, MRI and US are not routinely used for spleen size assessment in clinical practice (see challenges in Section \ref{Section 1.2}), where 2D US remains the standard imaging modality. Alternatively, computer vision methods have been proposed to estimate 3D shape from 2D images, bypassing the need for 3D US reconstruction. These include recent works like Pix2Vox++ \citep{xie2020pix2vox++}. Notably, Pix2Vox++ has been applied in medical imaging, specifically to 3D cardiac reconstruction from 2D US in \cite{stojanovski2022efficient}. However, reconstruction-based methods often require manual clinical landmarks to select salient 2D slices, and good reconstruction quality is normally only possible with multiple 2D views (more than two) acquired from different angles. In the spleen, it is impractical to obtain either multiple 2D images or high quality 3D US images due to the size of the spleen and the limited acoustic windows. This challenge has prevented us from adopting these methods.

\subsection{Variational Autoencoders in Medical Image Analysis}
VAEs are able to learn a low-dimensional representation (or latent space) from high-dimensional data such as images or segmentations, and provide benefits in terms of interpretability by allowing reconstruction to the original (image) space from their latent space. VAEs have been previously combined with a secondary task in a similar way to our framework. For example, \cite{biffi2018learning} proposed a VAE-based framework which also performed disease diagnosis from cine cardiac MRI derived segmentations. The use of a VAE enabled interpretability by visualising the learned features for healthy and hypertrophic cardiomyopathy patients. Similarly, in \cite{puyol2020interpretable}, a residual block-based VAE was trained using cine cardiac MRI segmentations to perform classification for predicting treatment response as well as explanatory concepts. In \cite{puyol2020assessing}, a VAE based on cardiac functional biomarkers was combined with a regression task in the latent space to learn the relationship between cardiac function and systolic blood pressure. \cite{cetin2023attri} proposed the Attri-VAE model to enable disentangled interpretation based on clinical attributes between images of healthy and myocardial infarction patients. In \cite{zhao2019variational}, a conditional generative VAE-based model was used to enable direct regression of brain age from 3D MRI images. \cite{joo2023variational} proposed a framework that utilised VAEs to learn features from various teeth images, and subsequently applied linear regression to estimate the corresponding subjects’ ages. To the best of our knowledge, VAEs have not yet been investigated for the direct estimation of 3D volume from 2D views. 

\subsection{Diffusion-based Generative Model}
In addition to developing a pipeline for spleen volume estimation, in this paper we also address the challenge of the lack of paired 2D spleen US images and corresponding 3D spleen segmentations for deriving accurate volume information. We address this problem by generating synthetic US data based on 3D CT data.

Traditional model-based simulators, such as Field II \citep{jensen1997field,jensen1992calculation} and SIMUS \citep{garcia2022simus}, have been developed to generate synthetic ultrasound images by modelling acoustic wave propagation and its interaction with models of tissues and organs. While these techniques can produce realistic ultrasound images, they require specific domain knowledge for hyperparameter tuning and must be applied case-by-case, which becomes computationally expensive for large-scale data synthesis. In addition, the quality of simulated ultrasound images is highly dependent on the fidelity of the underlying tissue or organ models, and they often fail to reproduce the artefacts caused by surrounding tissue or organs that interact with the primary target structure.

Because of these limitations, research focus has turned to the use of deep generative models for US simulation. Deep generative models provide a powerful and effective technique for generating high-quality synthetic images, which enables end-to-end learning for the images' patterns from large-scale datasets. In addition, deep generative models enable fast generation of diverse outputs from a single condition input simply by sampling from a latent distribution. Previous works based on generative adversarial networks (GANs) and Cycle GANs have shown the feasibility of deep adversarial models in generating realistic medical images \citep{armanious2019unsupervised,gilbert2021generating}. However, GANs are known to be highly unstable during training \citep{mescheder2018training} and suffer from a number of other drawbacks. These include the complexity and difficulty in achieving consistent descent with non-convex loss functions \citep{kodali2017convergence} and vanishing gradients \citep{saxena2021generative}. For these reasons, the generalisability of GANs proposed so far has been limited, and it is challenging for GANs to produce consistent and diverse synthetic data.

Denoising diffusion probabilistic models (DDPMs) are a more recent type of generative model based on maximum likelihood learning \citep{ho2020denoising,wang2022semantic}. Compared to GANs, the training process of DDPMs is more stable, and the generation is more diverse. Subsequently proposed conditional DDPMs, such as \cite{saharia2022palette,saharia2022image}, introduced the concatenation of condition information into the DDPMs, forcing the network to generate samples strictly according to the given condition information. For the medical image domain, the high variability between individual patients and the natures of different imaging modalities make the generation of synthetic medical images particularly challenging. Conditional DDPMs offer a potential solution to these challenges as they can generate synthetic images based on a provided anatomical structure (i.e., the condition). Despite the recent success of DDPMs in the domain of natural images, their application in medical image synthesis remains as an emerging area of research \citep{kazerouni2023diffusion}. For instance, \cite{lyu2022conversion} proposed a diffusion-based network to convert CT to MRI images, while \cite{pinaya2022brain} explored brain CT image generation. DDPMs have also been investigated for synthetic US image generation, including brain US \citep{pinaya2022brain} and musculoskeletal US \citep{katakis2023generation}. \cite{stojanovski2023echo} employed a layout consisting of cardiac segmentations and US cones as conditions for the denoising process and then evaluated the synthesis outcome by training a U-Net segmentation model using the synthetic data.

\section{Methods}
\label{Section 3}
\begin{figure*}[!htb] 
\centering    
\includegraphics[width=1\textwidth]{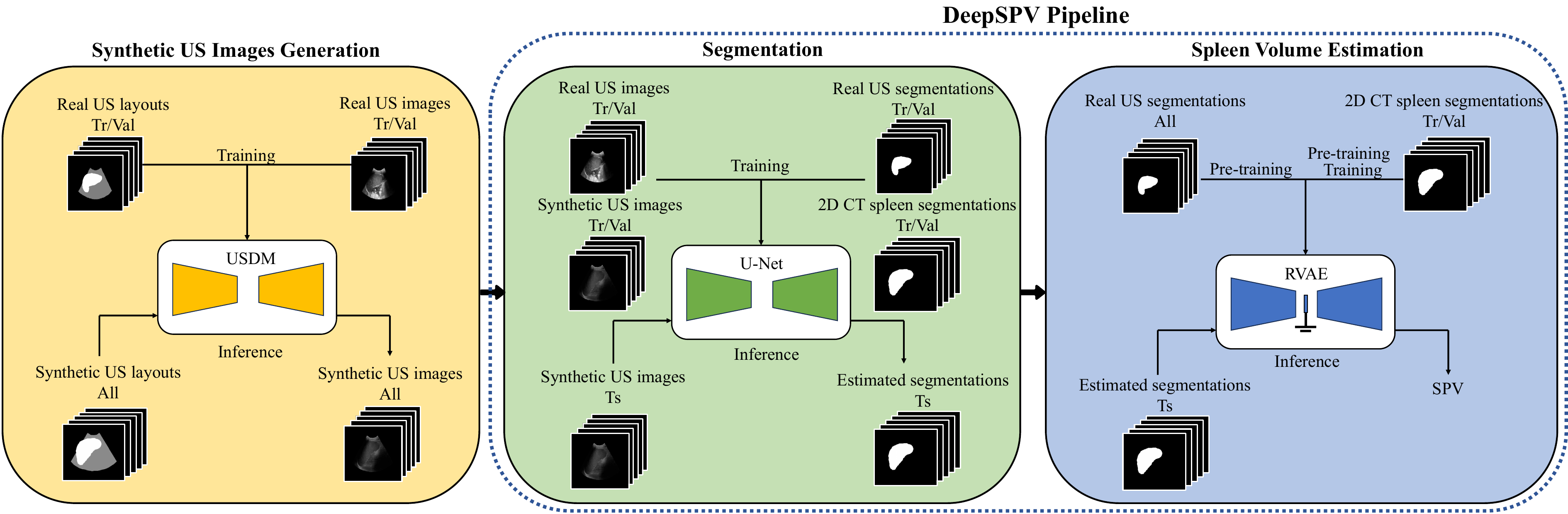}
\caption{An overview of this study. It first involves generating synthetic US images with the USDM, which is necessary due to the lack of paired 2D US images and ground truth volumes. The complete DeepSPV pipeline, proposed to estimate spleen volume from 2D US images, is enclosed within the blue dotted frame and consists of two parts: segmentation and spleen volume estimation. USDM: ultrasound semantic diffusion model. RVAE: end-to-end regression VAE. SPV: spleen volume. Tr/Val/Ts: training/valiation/test.}
\label{Figure 3.1}
\end{figure*}

In this section, we present the details of our DeepSPV for automated estimation of volume from 2D single- or dual-view spleen US images. Fig. \ref{Figure 3.1} provides an overview of the DeepSPV pipeline (see blue dotted frame) and the generation of synthetic US data for pipeline evaluation. The pipeline consists of a deep learning-based spleen segmentation model followed by a VAE-based volume estimation model.
Note that due to the lack of paired 2D US images and corresponding ground truth volumes, we utilised 3D manual spleen segmentations from a CT dataset to develop and evaluate our proposed methods. Details of this dataset are provided in Section \ref{Section 4.1.1}.

We start by providing a brief description of the 2D US image segmentation model in Section \ref{Section 3.1}. Then, we describe the VAE-based model for automated estimation of volume from single- or dual-view 2D spleen segmentations, in which we propose three different methods for volume estimation. Finally, Section \ref{Section 3.3} details our USDM, which is used to generate a synthetic US dataset for developing and evaluating the pipeline.

\subsection{2D Spleen US Image Segmentation}
\label{Section 3.1}
As described in Section \ref{Section 3}, our pipeline decomposes the process of spleen volume estimation from 2D US into two steps: automated spleen segmentation from US image(s) and the subsequent volume estimation from the predicted segmentation(s). In our previous work \citep{yuan2022deep}, we proposed the use of a U-Net for automated 2D spleen segmentation and length measurement, and obtained satisfactory segmentation accuracy. Therefore, in the proposed pipeline, we again employed a U-Net for the segmentation task. We used a standard 8-layer U-Net, which consists of 4 downsampling blocks, 1 bottleneck block, and 4 upsampling blocks. Each block involves three convolutional layers (each convolution followed by a batch normalisation and a ReLU activation function) and a max pooling or up-convolutional layer. 

\subsection{VAE-based Framework for 3D Volume Estimation from 2D Segmentations}
\label{Section 3.2}
After obtaining the automated 2D spleen segmentations from US images using the U-Net,  we propose a VAE-based framework to make a volume estimation from the estimated segmentation(s). 

We first introduce the architecture of the VAE in Section \ref{Section 3.2.1} and then describe three different methods for automatic volume estimation from the VAE latent space in Section \ref{Section 3.2.2}. Subsequently, in Section \ref{Section 3.2.3}, we provide details of an adapted framework that harnesses the generative nature of the VAE to provide estimated volumes with associated confidence intervals, which is aimed at improving the model’s utility for clinical end-users. 

\subsubsection{VAE Network Structure}
\label{Section 3.2.1}
\begin{figure*}[!htb] 
\centering    
\includegraphics[width=0.9\textwidth]{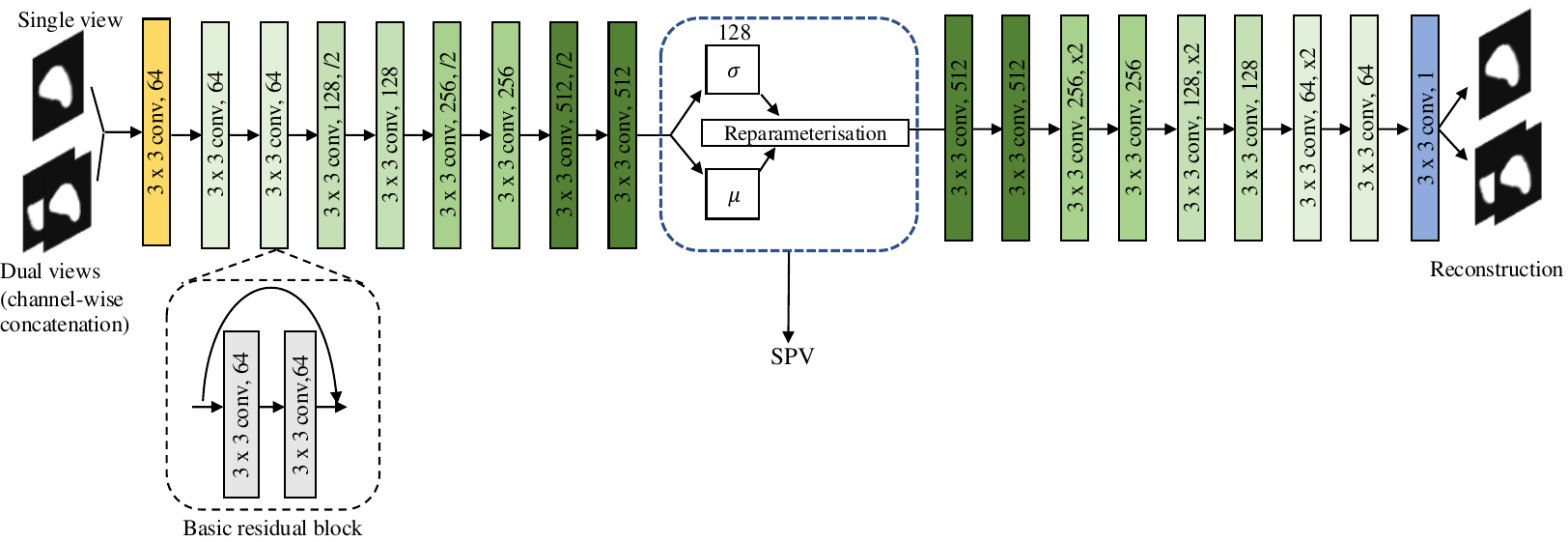}
\caption[An illustration of the proposed VAE-based framework.]{An illustration of the proposed VAE-based framework. The latent space distribution is parameterised by the mean $\mu$ and standard deviation $\sigma$
, which is shown in the blue dotted frame. The spleen volume is estimated from this distribution. The basic residual block is shown in the black dotted frame. All our VAE-based models were trained with either a coronal slice (single view) or a channel-wise concatenation of a coronal slice and a transverse slice (dual views). SPV: spleen volume.}
\label{Figure 3.2}
\end{figure*}

We employ a residual block-based convolutional VAE. The VAE consists of an encoder and a decoder, where the encoder compresses the input into a latent space distribution and the decoder restores the data from the low-dimensional representation back to the input space. The architecture is shown in Fig. \ref{Figure 3.2}. 

The encoder and decoder both consist of 8 basic residual blocks, each of which comprise two cascades of convolutional layers, batch normalisation, and ReLU activation function (see the detail of the basic residual blocks in Fig. \ref{Figure 3.2}). Input and output layers were used to process the input segmentations and output features. The mean $\mu$ and standard deviation $\sigma$ parameterise the latent space distribution, each with a dimension of 128. During training, the encoder learns to map the inputs to the latent space by sampling based on the $\mu$ and the $\sigma$, and the decoder restores the segmentation using the reparametrised representation $z=\mu +\zeta \bigodot \sigma $, where $\zeta \sim N\left ( 0,I \right )$ is a noise vector and $\bigodot$ represents element-wise multiplication. 

The loss function used to optimise the VAE comprised reconstruction loss and regularisation loss. The reconstruction loss is measured using binary cross entropy (BCE) that is calculated between the original label $\mathbf{s}$ and the reconstructed label $\hat{\mathbf{s}}$. The regularisation loss is calculated through the Kullback-Leibler divergence (KLD) to regularise the latent distribution weighted by $w_{1}$. The BCE loss is defined in Equation \ref{Equation 3.1}:

\begin{equation}
    \begin{split}
    \textit{BCE}(\mathbf{s}_i, \hat{\mathbf{s}}_i) = -\frac{1}{HW} \sum_{j=1}^{H} \sum_{k=1}^{W} \Big[ \mathbf{s}_{i,jk} \log \hat{\mathbf{s}}_{i,jk} \\
    + (1 - \mathbf{s}_{i,jk}) \log (1 - \hat{\mathbf{s}}_{i,jk}) \Big]
    \end{split}
    \label{Equation 3.1}
\end{equation}

\noindent where $\mathbf{s}_i$ and $\hat{\mathbf{s}}_i \in \mathbb{R}^{H \times W}$ are the input and reconstructed segmentations for the $i$-th sample, and $\mathbf{s}_{i,jk}$ and $\hat{\mathbf{s}}_{i,jk}$ are their pixel values at location $(j,k)$. $H$ and $W$ represent the height and width of the images.

The KLD loss is presented in Equation \ref{Equation 3.2}:

\begin{equation}
    \textit{KLD}(\mu_i, \sigma_i) = -\frac{1}{2} \sum_{d=1}^{D} (1 + \log \sigma_{i,d}^{2} - \mu_{i,d}^{2} - \sigma_{i,d}^{2})
    \label{Equation 3.2}
\end{equation}

\noindent where $\mu_i$ and $\sigma_i$ are the mean and standard deviation for the $d$-th dimension of the latent representation for the $i$-th sample, and $D$ is the total dimension of the latent representation. In our case, $D = 128$.

Finally, the overall loss is defined below in Equation \ref{Equation 3.3}:

\begin{equation}
    \mathcal{L} = \frac{1}{N} \sum_{i=1}^{N} \Big[ \textit{BCE}(\mathbf{s}_i, \hat{\mathbf{s}}_i) + w_{1} \cdot \textit{KLD}(\mu_i, \sigma_i) \Big]
    \label{Equation 3.3}
\end{equation}

\noindent where $N$ is the size of a mini-batch, and $w_1$ is a weighting term for the KLD loss.

Note that all models were trained in two settings – with only a coronal slice (single view) and also with a channel-wise concatenation of coronal and transverse slices (dual views). See Section \ref{Section 4} for experimental details. 

\subsubsection{Spleen Volume Estimation}
\label{Section 3.2.2}
We propose and investigate three different methods to estimate spleen volume from the latent space distribution of the convolutional VAE. Each of these is described below. \\

\begin{figure*}[!htb] 
\centering    
\includegraphics[width=0.8\textwidth]{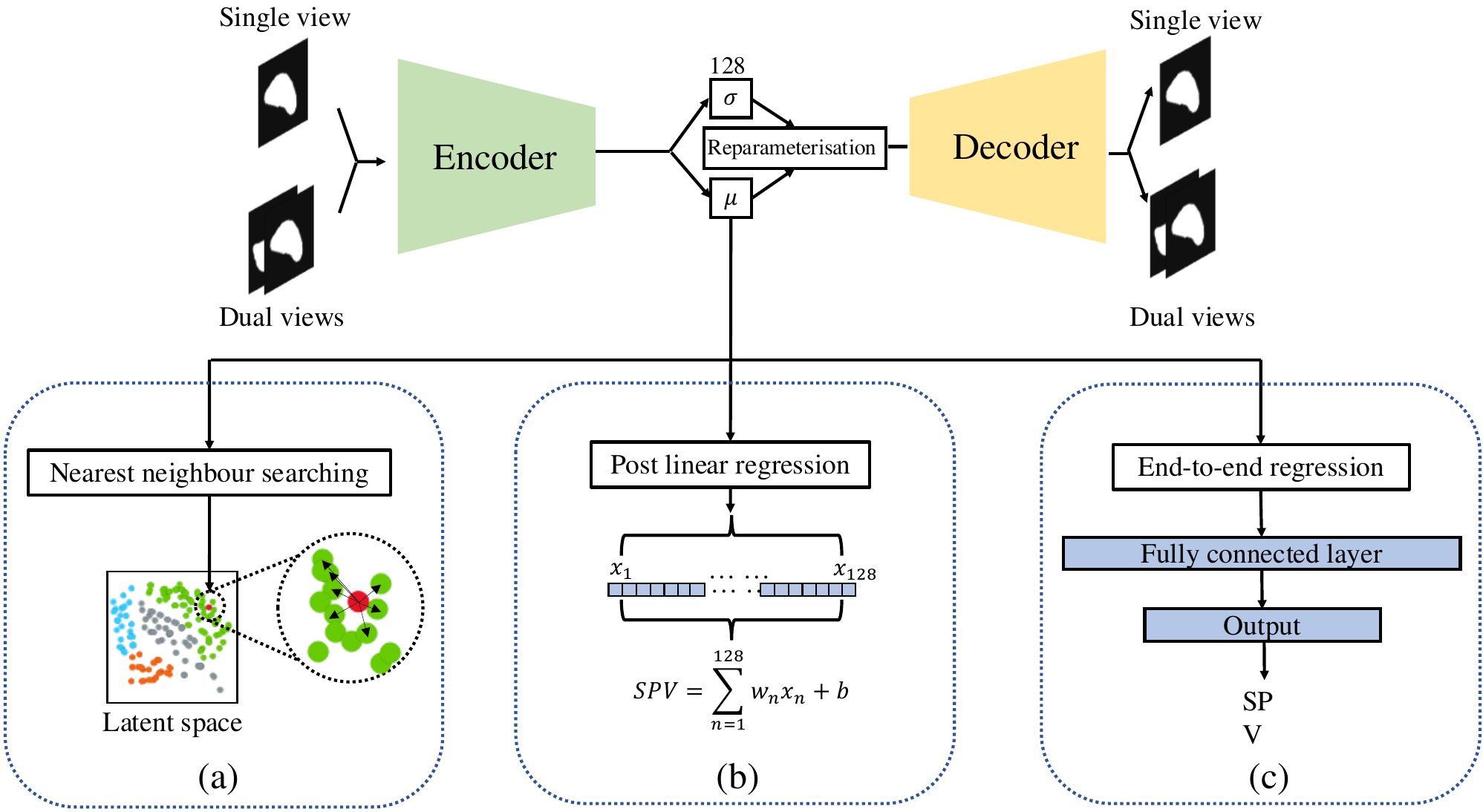}
\caption{Illustration of three proposed VAE-based methods: (a) Nearest neighbour searching in the latent space (NN). (b) Post linear regression of latent representations (PLR). (c) End-to-end regression VAE (RVAE).}
\label{Figure 3.3}
\end{figure*}

\noindent\textbf{Nearest neighbour searching in the latent space (NN):} In this method, we first trained the VAE to model the distribution of observed segmentations. Then, during inference, we mapped the input segmentation into the latent space and searched for the nearest neighbour from the training dataset. The corresponding ground truth volume value of this nearest neighbour was used as the volume estimate for the test case. 

Formally, we denote by $\mathbf{s}$ the input image(s), which can be a single coronal slice, or a channel-wise concatenation of a coronal slice and a transverse slice. Superscripts $Tr$ and $Ts$ are used to indicate whether slices are from the training or test set. During inference, we inputted $\mathbf{s}^{Ts}$
into the VAE, resulting in the latent space representation mean $\mu ^{Ts}$, whose dimension is 128. We then inputted each sample in the training dataset $\mathrm{S}^{Tr}=\left \{ \mathbf{s}_{1}^{Tr},\mathbf{s}_{2}^{Tr},\mathbf{s}_{3}^{Tr},...,\mathbf{s}_{N}^{Tr} \right \}$ into the trained VAE and get representations $\mathrm{M}^{T}=\left \{ \mu_{1}^{Tr},\mu_{2}^{Tr},\mu_{3}^{Tr},...,\mu_{N}^{Tr} \right \}$. We calculated the Euclidean distance between each $\mu _{n}^{Tr} $ and $\mu ^{Ts}$, and used the closest training sample’s corresponding spleen volume as the estimated volume value for the test subject.

In this method we used the basic loss function presented in Equation \ref{Equation 3.3} to train the VAE. An illustration of this volume estimation approach is shown in Fig. \ref{Figure 3.3}a. \\

\noindent\textbf{Post linear regression of latent representations (PLR):} In this method, after the VAE was trained using the basic loss function presented in Equation \ref{Equation 3.3}, a simple linear regression was performed on the $\mu$ of the training data to predict the spleen volume. At inference time, each input was embedded into the latent space and its $\mu$ value used with the obtained regression coefficients to estimate the volume. Fig. \ref{Figure 3.3}b illustrates the proposed PLR method. \\

\noindent\textbf{End-to-end regression VAE (RVAE):} This method was also based upon regression of volume from the VAE latent space, but here we added additional network layers to the VAE latent space to perform the regression. The weights of the fully connected layer were optimised to predict the spleen volume from the latent space in an end-to-end training manner, enabling modelling of the potentially non-linear relationship between the latent embeddings and spleen volume. Specifically, we added a fully connected layer with size 64 and ReLU activation function, followed by an output layer to predict the volume from the features. For RVAE, during training, we included an extra regression term to the basic loss function in Equation \ref{Equation 3.4} to train the network to learn to estimate the spleen volume. 

\begin{equation}
    \begin{split}
    \mathcal{L} = \frac{1}{N} \sum_{i=1}^{N} \Big[ \textit{BCE}(\mathbf{s}_i, \hat{\mathbf{s}}_i) + w_{1} \cdot \textit{KLD}(\mu_i, \sigma_i) \\
    + w_{2} \cdot \textit{MSE}(vol_i, \hat{vol}) \Big]
    \end{split}
    \label{Equation 3.4}
\end{equation}

\noindent where $vol_i$ is the ground truth volume value, $\hat{vol}_i$ is the estimated volume value, and $w_1$ and $w_2$ are weights for the KLD loss and mean squared error (MSE) losses. An illustration of the RVAE method is shown in Fig. \ref{Figure 3.3}c. 

\subsubsection{Volume Estimation with Confidence Intervals}
\label{Section 3.2.3}
To further harness the generative nature of VAEs, we also integrated a method to compute 95\% confidence intervals in addition to a scalar estimated volume. Such information could be beneficial to clinicians in evaluating possible splenomegaly using US and allow greater flexibility in the use of our method as a decision support tool. We implemented this approach only for the RVAE method due to its superior volume estimation performance (see Section \ref{Section 5.2.1}). In order to train the RVAE network to make multiple predictions for a single input, we modified the input to the fully connected layers $\mu$ to the reparametrised representation $z=\mu +\zeta \bigodot \sigma $ in the training stage, which is the same as the input provided to the decoder of the VAE. At inference time, we sampled in the latent space 100 times using $\zeta \sim N\left ( 0,I \right )$. For each sampled reparametrised representation $z$, we inputted it into the fully connected layers to estimate the spleen volume using the method described above. The 95\% confidence interval was then determined by $\eta \pm 1.96\times \theta$, where $\eta$ and $\theta$ are the mean and standard deviation of the 100 volume estimates. We denote this method by RVAE-CI. 

\subsection{Synthetic US Image Generation Using USDM}
\label{Section 3.3}
In this section, we provide details of the proposed USDM for synthetic but realistic US image generation from CT derived segmentations. Note that synthetic US images were generated only for the single-view (coronal) condition. The transverse view was not synthesised, as we only have access to coronal real 2D US images, which are typically recorded during spleen length assessment in clinical practice.

\begin{figure}[!htb] 
\centering    
\includegraphics[width=0.48\textwidth]{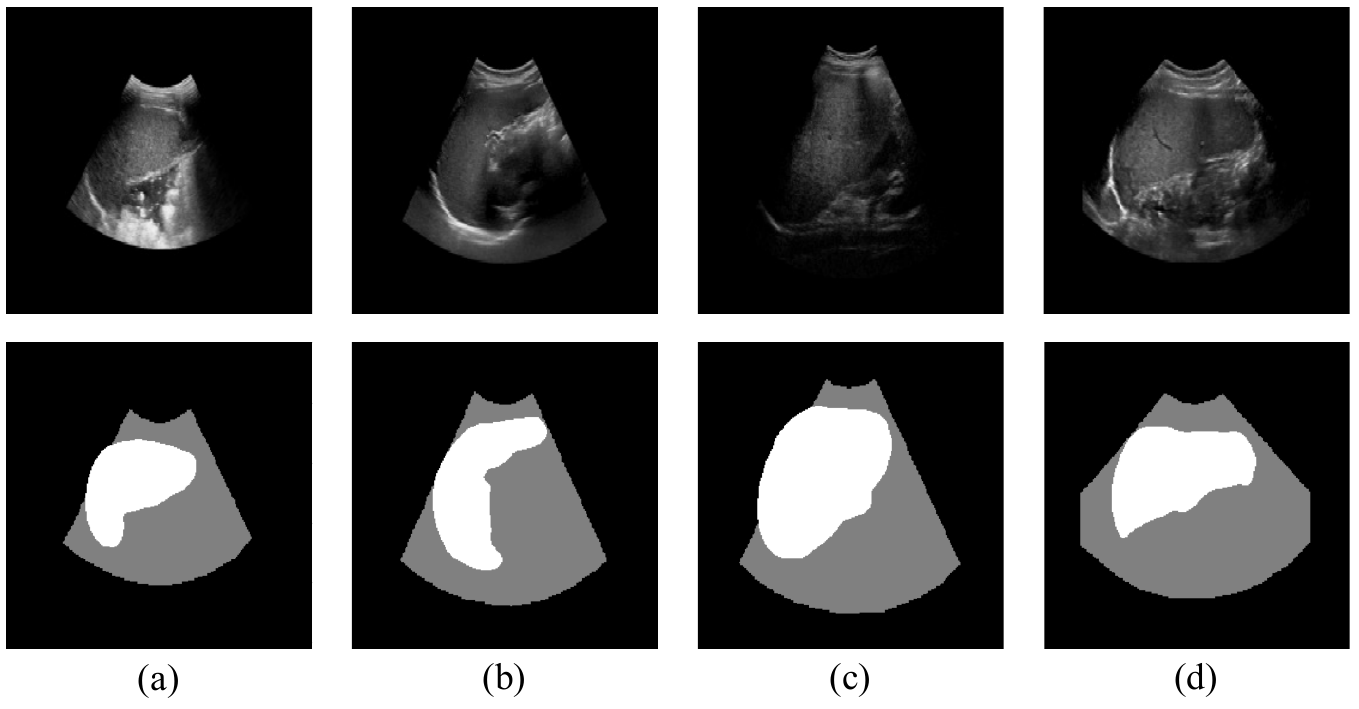}
\caption{US images and their corresponding US layouts. The top row displays the original US images while the bottom row presents their corresponding US layouts. Each column refers to a specific case.}
\label{Figure 3.4}
\end{figure}

\begin{figure}[!htb] 
\centering    
\includegraphics[width=0.48\textwidth]{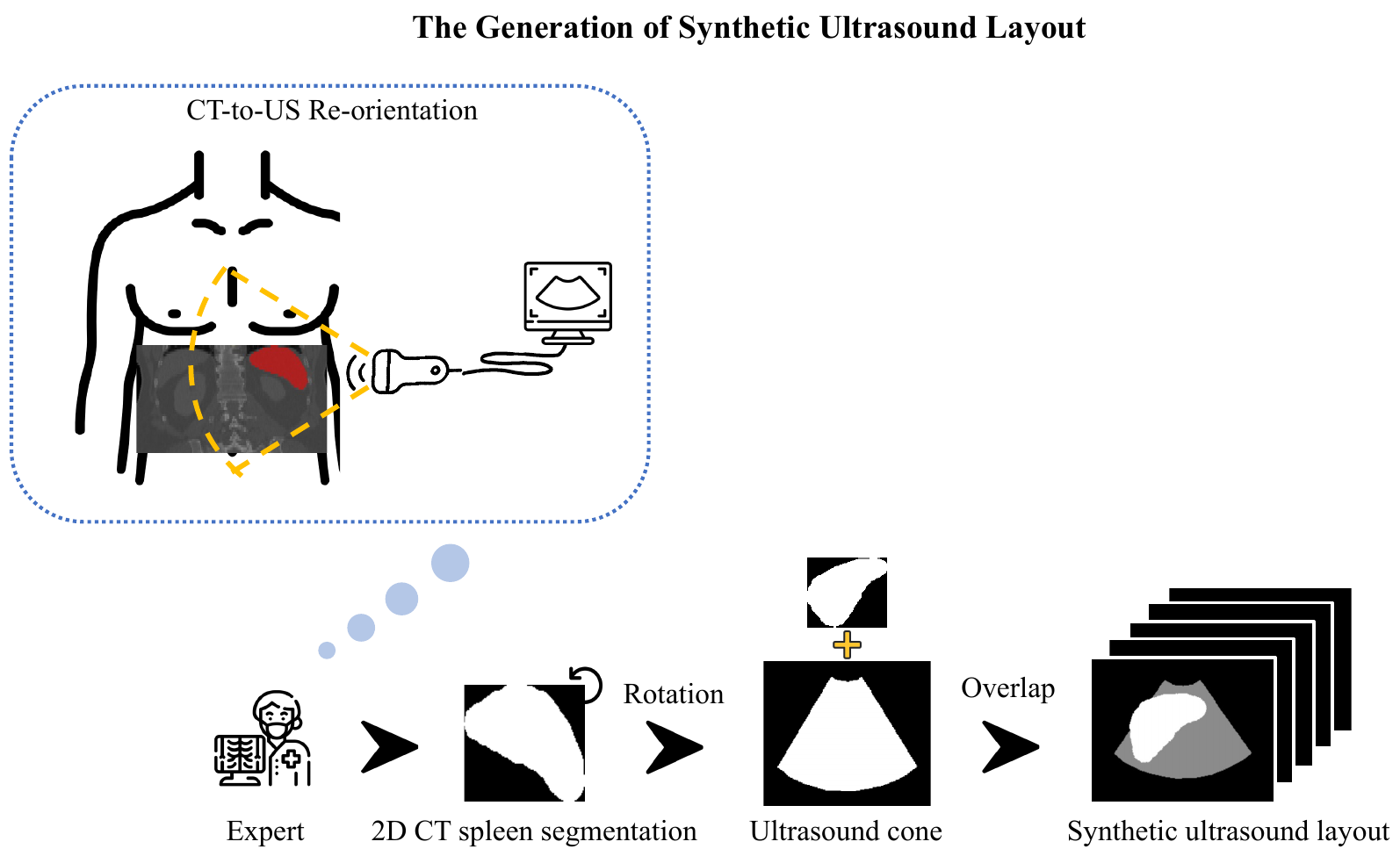}
\caption{Illustration of the synthetic US layout creation process. The diagram shows how 2D CT spleen segmentations were rotated and aligned with real US cones under expert guidance to emulate realistic clinical acquisition. Specifically, to match the typical left-sided relative location of US probes to the patient in a US examination, the spleen segmentation from CT was rotated 90 degrees anti-clockwise and manually overlaid with an appropriately sized US cone by a human expert.}
\label{Figure 3.5}
\end{figure}

We begin by detailing the manual creation of the spleen US `layout' images. These layout images specify the extent of the US cone and the ground truth spleen segmentation in 2D. Examples of these layout images are provided in Fig. \ref{Figure 3.4} and Fig. \ref{Figure 3.5} illustrates the way in which they were produced. The US layouts for training the USDM were formed from real US images and ground truth segmentations of the spleen and cone. The US layouts used at inference were formed by combining 2D CT spleen segmentations and selected real US cones. For each generation of a synthetic US layout, we manually rotated a 2D spleen segmentation and aligned it with a real US cone of an appropriate size. To be specific, as shown in Fig. \ref{Figure 3.5}, the ultrasound probe is typically placed on the left-side ribs of the patient. To align the relative spatial location of the spleen in CT with that in the 2D US images, we rotated the spleen segmentation from the CT 90 degrees anti-clockwise. This ensures that the relative orientation of the spleen matches that in the clinical US acquisition. The rotated spleen segmentation was then manually placed on a real US cone by a human expert.

\begin{figure*}[!htb] 
\centering    
\includegraphics[width=0.9\textwidth]{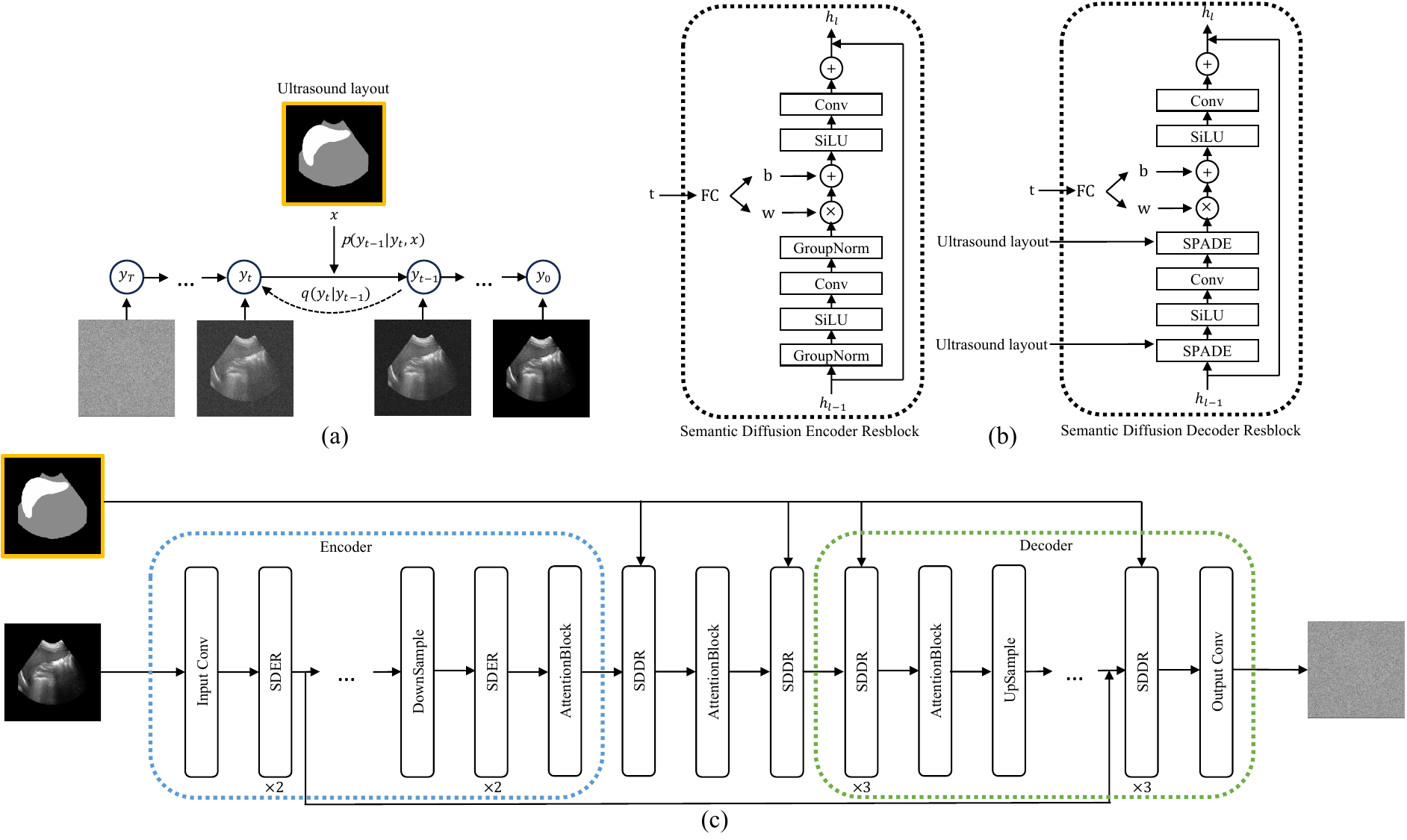}
\caption{(a) US Semantic Diffusion Model (USDM). The framework transforms the noise from a standard Gaussian distribution to a realistic image through an iterative denoising process guided by the US layout $x$. (b) Semantic Diffusion Encoder Resblock (SDER) and Semantic Diffusion Decoder Resblock (SDDR). (c) The detailed architecture of the USDM.}
\label{Figure 3.6}
\end{figure*}

In a USDM, two essential processes are defined, the forward process and the reverse process. In the forward process, the original data is gradually made more noisy in scheduled steps $t\epsilon \left [ 0,T \right ]$. At each scheduled step $t$, the data is perturbed by Gaussian noise $y_{t}=y_{t-1}+q\left ( y_{t}|y_{t-1} \right )$, where $y_{t}$ represents the noisy data after the current step, and $q\left ( y_{t}|y_{t-1} \right )$ denotes the added noise. Conversely, in the reverse process, the model learns to estimate the noise at the current step with the guiding condition $x$ from the noisy data $y_{t} $:$p\left ( y_{t-1}|y_{t},x \right )$. Fig. \ref{Figure 3.6}a gives an overview of our USDM conditioned by the spleen US layouts. We adopted a similar framework to \cite{wang2022semantic} as the backbone of our network. While the network is built based on a U-Net architecture, it features several modifications. The semantic diffusion encoder resblock (SDER) and semantic diffusion decoder resblock (SDDR) are introduced to improve the performance of the synthetic generation. Fig. \ref{Figure 3.6}a-\ref{Figure 3.6}c illustrates the denoising process and the details of the USDM. 

\section{Experiments}
\label{Section 4}
\subsection{Materials}
\label{Section 4.1}
In this study, we utilised a real 2D US dataset with ground truth spleen segmentations and 3D spleen segmentations derived from a CT dataset. The descriptions of these datasets are provided in Section \ref{Section 4.1.1}. Details of the data pre-processing process is given in Section \ref{Section 4.1.2}.

\subsubsection{Dataset Description}
\label{Section 4.1.1}
The CT segmentation dataset is comprised of 149 manual segmentations of CT volumes from two publicly available collections, with 36 exhibiting splenomegaly (characterised by a volume greater than 314.5mL) and the remaining 113 having a volume that falls within a normal range \citep{linguraru2013assessing,bezerra2005determination,de1999normal}. 60 spleen segmentations were accessed from the Medical Segmentation Decathlon (MSD) challenge \citep{antonelli2022medical}. This dataset consists of 40 3D CT volumes with associated manual segmentations (41 cases in total, but one image lacks part of the spleen and was therefore excluded from our experiments) and 20 CT volumes without manual segmentations. An experienced radiologist performed additional manual segmentations of these 20 CT volumes, resulting in a final dataset of 60 manual segmentations of CT volumes from MSD. The second data source is from \cite{gibson2018automatic}, which consists of 90 CT volumes with manual segmentations of multiple abdominal organs including the spleen. This dataset is a combination of 43 subjects from the Cancer Imaging Archive Pancreas-CT dataset \citep{roth2015deeporgan,clark2013cancer} and 47 subjects from the ‘Beyond the Cranial Vault’ (BTCV) segmentation challenge \citep{landman2016}. One BTCV segmentation was excluded due to an excessively large spleen volume (3083mL), which was considered an outlier even for a case of splenomegaly, as the average spleen volume for a splenomegaly case has been determined to be 1004.75 ± 644.27mL \citep{linguraru2013assessing}. Thus, in total, 89 segmentations from Gibson et al. were utilised in our study, and all manual spleen segmentations were obtained directly from the publishers. Note that for all datasets only the corresponding spleen segmentations of CT volumes were utilised in our work, not the original CT volumes themselves. 

For the 2D US dataset, all images were obtained from paediatric patients with SCD, who were under professional disease management at Evelina London Children's Hospital at Guy's and St Thomas' NHS Foundation Trust from 2005 to 2020. All images were anonymised, and patient information was removed from the images before use. The use of this dataset for research purposes has been ethically approved under clinical audit project number 9028 by Guy's and St Thomas' NHS Foundation Trust on 12 November 2018. The patients underwent multiple US examinations as part of their routine clinical monitoring, and for each examination, 1 US image was recorded for spleen length measurement as part of routine clinical management. Typically, for measuring spleen length, a coronal or nearly coronal US view with the longest spleen length was recorded. The dataset contains up to 4 images for each patient, each acquired at an examination that occurred at a different time point. In total 363 US images were retrieved. The spleen and US cone were manually segmented by a trained observer in all of these images. Note that the US cones also required manual segmentation, as acquisition parameters were unavailable in our retrospectively retrieved dataset.

\subsubsection{Data Pre-processing}
\label{Section 4.1.2}
Given the differences in image size and resolution between the CT and US datasets, it is crucial to ensure consistency between the two modalities’ images so that they can be combined for model training and validation. Our first step was spatial alignment. All 3D CT spleen segmentation volumes were resampled to a voxel size $1\times 1 \times 1{mm}^{3}$. Similarly, the US images and their corresponding spleen segmentations were also re-sampled to a pixel size of $1\times 1{mm}^{2}$. 

\begin{figure}[!htb] 
\centering    
\includegraphics[width=0.48\textwidth]{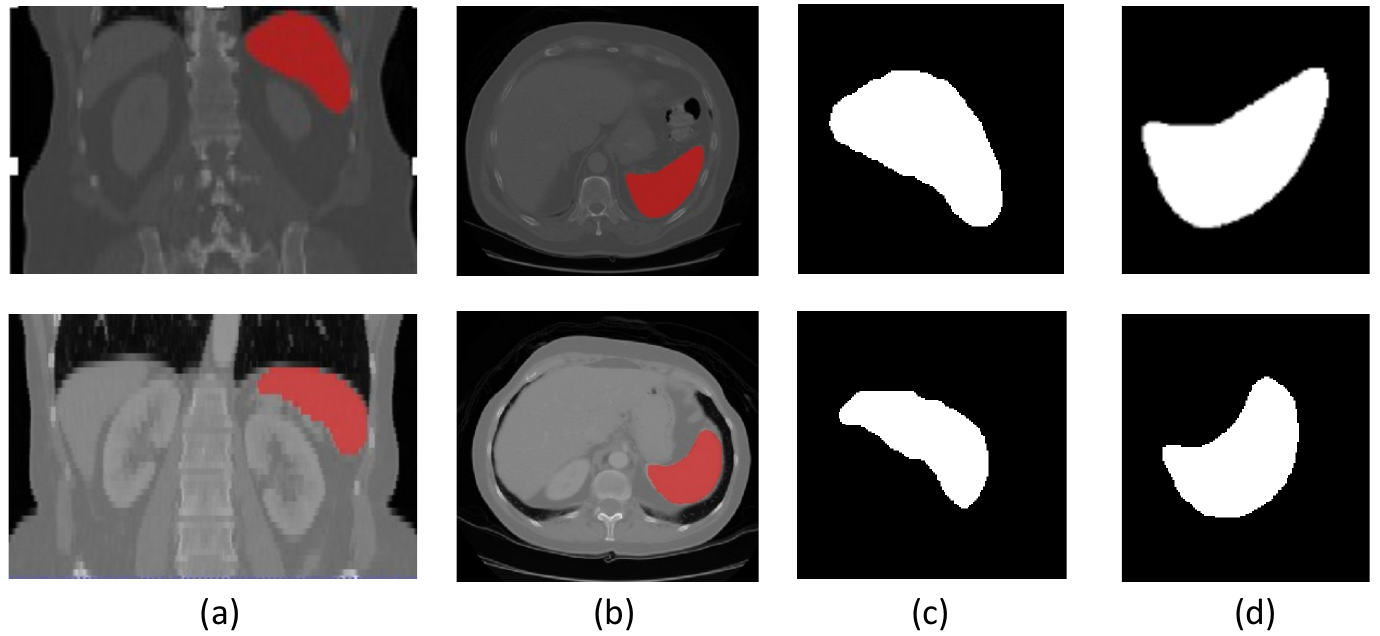}
\caption{Two example CT volumes (top and bottom rows) with associated ground truth spleen segmentations and the selected coronal and transverse 2D segmentation slices. (a) Coronal view of CT volume with spleen segmentation in red. (b) Transverse view of CT volume with spleen segmentation in red. (c) Selected coronal 2D segmentation slice with largest cross-sectional area. (d) Selected transverse 2D segmentation slice with largest cross-sectional area.}
\label{Figure 4.1}
\end{figure}

For the VAE-based volume estimation experiment,  the 3D spleen segmentations were cropped based on calculated bounding boxes and centroids. Specifically, we translated the segmentations so that their centroids aligned, followed by padding to a size of $164\times 186\times 176$ to maintain consistency according to the largest bounding box across the dataset. We then selected the 2D coronal slice and transverse slice with the largest spleen cross-sectional area as experimental data from each 3D spleen segmentation (see Fig. \ref{Figure 4.1}). All 2D slices were then resized to a size of $256\times 256$.

For the full pipeline evaluation, the US images/segmentations and coronal slices selected from the CT segmentations were all padded to a size of $190 \times 272$ to match the size of the bounding box of the largest US cone. Finally, all $190\times 272$ spleen segmentations were resized to a size of $256\times 256$. Note that all images used for training the USDM and the U-Net were either cropped or zero-padded to match the height and width of the largest US cone in the dataset, and all the 2D spleen segmentations employed to train the VAE were centre-cropped based on the spleen segmentations.  

\subsection{Implementation and Experimental Details}
\label{Section 4.2}
\subsubsection{Volume Estimation and Confidence Intervals}
\label{Section 4.2.1}
As explained in Section \ref{Section 4.1.1}, we employed 149 segmentation volumes from two different sources to develop and evaluate our proposed volume estimation methods. For each 3D spleen segmentation, we generated two types of 2D segmentations – one with only a coronal slice (single view) and the other with the channel-wise concatenation of a coronal slice and a transverse slice (dual views). For each proposed volume estimation method, we trained models using these 2D segmentations under the single-view and dual-view scenarios, resulting in two sets of results for each of the proposed models.  

To evaluate the performance of the models, we split the dataset into 5 folds, with each fold containing 30 segmentations. Two folds contained a duplicated case, but it was not included in the hold-out test fold. Each fold had 7-8 segmentations manifesting splenomegaly (to be more specific, 7 for four folds and 8 for one fold), i.e., its volume was greater than 314.5mL. One fold (30 segmentations including 7 splenomegaly cases) was regarded as the hold-out test set, and to analyse the proposed methods comprehensively, we conducted a 4-fold cross-validation on the remaining folds, with 3 folds used for training and 1 fold used for validation. The evaluation of each method was based on the average performance on the hold-out test set of the 4 models trained from this cross-validation.  

We employed different training strategies for different methods. Specifically, for NN and PLR, we trained the model for up to 500 epochs with $w_{1}$ set to 0.2 (refer to Equation \ref{Equation 3.3}). For the RVAE-based methods, i.e., RVAE, and RVAE-CI, we first trained models using only the VAE loss (i.e., binary cross entropy and the Kullback-Leibler divergence loss) with $w_{1}=0.2$ and $w_{2}=0$ for 150 epochs (see Equation \ref{Equation 3.3}), and then trained them with both VAE loss and regression loss ($w_{1}=0.2$ and $w_{2}=0.2$) for up to 650 epochs, see Equation \ref{Equation 3.4}. For each evaluated method and each cross-validation fold, the optimal values of the hyperparameters $w_{1}$ and $w_{2}$ were found with a grid search strategy for values from 0.1 to 0.5 in steps of 0.1 using the validation set performance as the selection criterion. We also conducted a grid search over learning rate (values 0.0001, 0.001 and 0.01) and minibatch size (values 4 and 8). Based on this, we set learning rate = 0.001 and minibatch size = 8 for all methods. 

During the training phase of all methods, we performed data augmentation by applying small random rotations ranging from -15 to 15-degrees along all three axes to the 3D segmentations prior to the selection of the coronal and transverse slices. We saved the final trained models as those with the best volume estimation performance on the validation set over all epochs. All models were trained on a NVIDIA GeForce GTX TITAN X 24GB using the Adam optimiser. Note that the ground truth volume values were scaled down by a factor of 10 for use in training/inference, and the estimated volume values were scaled up by a factor of 10 after inference. This was done to achieve better convergence during training.

\textbf{Comparative Evaluation:} For comparative evaluation, we evaluated the Pix2Vox++ 2D-to-3D reconstruction framework \citep{xie2020pix2vox++}. Rather than estimating volume directly, the Pix2Vox++ method estimates a 3D segmentation, from which we computed the volume. Pix2Vox++ includes two versions: a version with lower computational complexity called Pix2Vox++/F and an advanced version with higher computational complexity called Pix2Vox++/A. We evaluated both versions on our data. 

The same data stratification and cross-validation approaches were applied for the comparative approaches. The Pix2Vox++ models were trained to reconstruct 3D segmentations from both single-view data and dual-view data, and we obtained the spleen volume by the product of the number of the voxels within the predicted 3D spleen and the voxel size. Following the process in the original paper, we resized the 3D ground truth volume to $32\times 32\times 32$ for use in training/inference. The estimated volume was obtained by rescaling the reconstructed 3D segmentations from $32\times 32\times 32$ back to $164\times 186\times 176$. We trained both models with a learning rate of 0.001 (chosen from 0.0001, 0.001 and 0.01) and a minibatch size of 8 (chosen between 4 and 8) after hyperparameter optimisation.  

Furthermore, as additional comparative approaches, we investigated several deep learning regression models: VGGNet \citep{simonyan2014very}, DenseNet \citep{huang2017densely} and ResNet \citep{he2016deep}. We chose VGG-16, ResNet-18, and DenseNet-121 due to their established performance and the comparable number of trainable parameters to our RVAE encoder (approximately 14.7M for VGG-16, 11.2M for both ResNet-18 and our RVAE encoder, and 6.9M for DenseNet-121, which is relatively lightweight among the three). We employed the same data preparation strategy as used for training our VAE based methods (including single- and dual-view settings, data size, augmentation, stratification, and cross-validation). Furthermore, we adopted the same fully connected layers as those in RVAE. We trained these regression models with MSE loss for up to 800 epochs (matching the total number of epochs used for training RVAE). For hyperparameter optimisation we tested learning rates of 0.0001, 0.001, and 0.01, and minibatch sizes of 4 and 8, and based on the results we set the learning rate to 0.001 and the minibatch size to 8 for all three comparative regression models.

Lastly, we also compared with current clinical methods for spleen volume estimation \citep{bezerra2005determination,prassopoulos1997determination}. These approaches are based on linear regression of volume from manual 2D measurements of spleen size made from cross-sectional views. The first variant used a single measurement of maximal spleen length $L$ based on  Equation \ref{Equation 4.1}, as proposed in \cite{bezerra2005determination},  

\begin{equation}
    \hat{vol}=\frac{\left ( L-5.8006 \right )}{0.0126}
    \label{Equation 4.1}
\end{equation}

\noindent and the second variant used three orthogonal measurements (spleen length $L$, maximal width $W$ and thickness at the spleen hilum \textit{Th}) based on Equation \ref{Equation 4.2}, as  proposed in \cite{prassopoulos1997determination}.  

\begin{equation}
    \hat{vol}=30+ 0.58\left ( W\times L\times \mbox{\textit{Th}} \right )
    \label{Equation 4.2}
\end{equation}

\begin{figure}[!htb] 
\centering    
\includegraphics[width=0.26\textwidth]{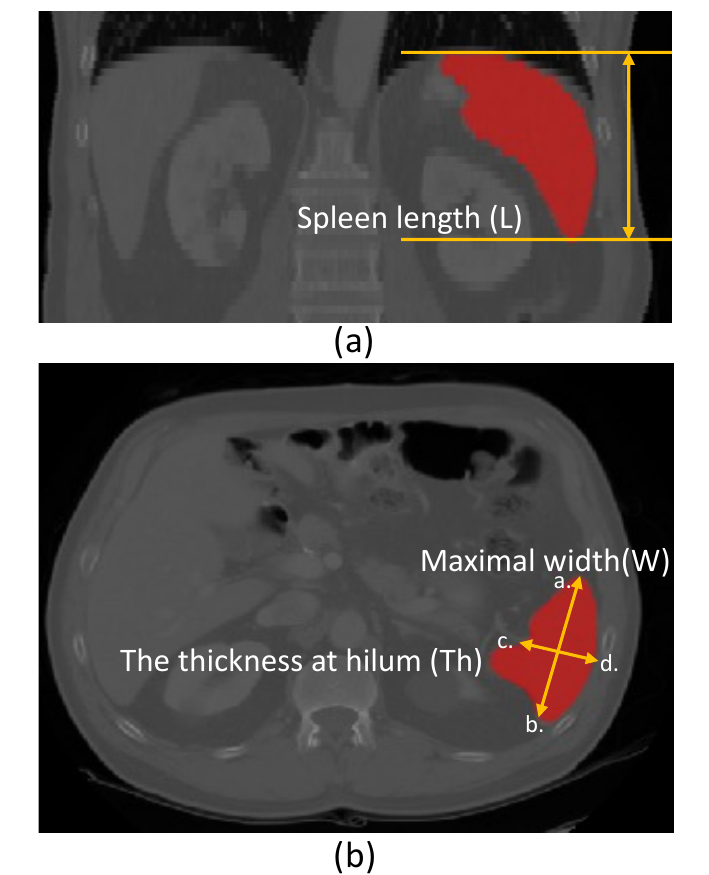}
\caption{Illustration of manual measurements from CT images for spleen volume estimation. (a) Coronal view. The spleen length (L) is obtained by multiplying the spacing of the coronal plane by the number of contiguous slices that contain the spleen along the transverse axis. (b) Transverse view. The maximal width (W) is obtained by finding the largest diameter on any transverse slice (see distance between ab). The thickness at hilum (Th) is determined by the thickness of the spleen at hilum and perpendicular to the spleen width (see distance between cd).}
\label{Figure 4.2}
\end{figure}

To obtain the spleen length, a human expert manually identified the most superior and inferior transverse slices that contained the spleen and multiplied the number of slices between the identified slices by the thickness of the transverse slices. The human expert also measured the maximal width and the thickness at hilum, where the maximal width of the spleen was defined as the largest diameter on any transverse slice and the thickness at hilum was determined by the thickness of the spleen at hilum and perpendicular to the spleen width. An example of the manual measurement process is illustrated in Fig. \ref{Figure 4.2}.

\subsubsection{USDM Training and Inference}
\label{Section 4.2.2}
We first trained the USDM model to generate coronal synthetic US images guided by the US layouts. During the training phase, we used the training and validation sets of the real US images/layouts. We conducted hyperparameter tuning on the annealed learning rate ($10^{-3}$, $10^{-4}$, and $10^{-5}$) and the minibatch size (4, 8, and 12). Based on the model's performance on the validation set of real US images/layouts and considerations of computational efficiency, we trained the USDM using an annealled learning rate of $10^{-4}$ and a minibatch size of 8. Both forward and reverse processes were defined with 1000 diffusion steps, and 100,000 complete denoising processes. Once the training was complete, we evaluated the USDM's performance on the test set of real US images/layouts to see how it performed with the real US layout as input. Subsequently, we employed all 149 synthetic US layouts (as detailed in Section \ref{Section 3.3}), including training, validation, and test sets, to generate 149 synthetic US images for the following segmentation and volume estimation experiments.

\subsubsection{Final DeepSPV Evaluation}
\label{Section 4.2.3}
We now describe the evaluation of the full DeepSPV pipeline using the synthetic US data. Note that for this evaluation we used only the RVAE volume estimation approach because of its superior performance as will be presented in Section \ref{Section 5.2.1}. In addition, due to the data limitation mentioned in Section \ref{Section 3.3}, the evaluation was performed only for estimation from single coronal 2D US images.\\

\noindent\textbf{Data Stratification:} We employed data from two modalities: US and CT. Specifically, we utilised three types of data: US images, spleen segmentations, and US layouts. We had a total of 363 US images, which were split into training (218), validation (72), and test (73). A total of 149 CT volumes were used, which were split into 89/30/30 for training/validation/test. The US layout images were derived from the CT and US images and were used to produce synthetic US images with known ground truth spleen volumes. A detailed breakdown of the data employed and their stratification into (pre-) training, validation and inference for the different parts of the pipeline is presented in Table \ref{Table S1}. Note that there is no data leakage between the development and evaluation of different models in the pipeline, i.e., no data used for training any part of the framework was used for testing purposes.\\

\begin{table*}[!htb]
\caption{The breakdown of the data used in the experiments to validate the full volume estimation pipeline. Tr: training. Val: validation. Ts: test. Pre-tr: pre-training. Inf: inference.}
\centering
\begin{tabular}{ccccccccc}
\hline
\textbf{} &
  \textbf{} &
  \multicolumn{2}{c}{\textbf{USDM}} &
  \multicolumn{2}{c}{\textbf{U-Net}} &
  \multicolumn{3}{c}{\textbf{RVAE}} \\ \hline
\textbf{} &
  \textbf{Data} &
  \textbf{Tr} &
  \textbf{Inf} &
  \textbf{Tr} &
  \textbf{Inf} &
  \textbf{Pre-tr} &
  \textbf{Tr} &
  \textbf{Inf} \\ \hline
\multirow{2}{*}{\textbf{Real US images}} &
  Tr/Val (218/72) &
  \checkmark &
  \ding{53} &
  \checkmark &
  \ding{53} &
  \ding{53} &
  \ding{53} &
  \ding{53} \\ \cline{2-9} 
 &
  Ts (73) &
  \ding{53} &
  \checkmark &
  \ding{53} &
  \checkmark &
  \ding{53} &
  \ding{53} &
  \ding{53} \\ \hline
\multirow{2}{*}{\textbf{Real US segmentations}} &
  Tr/Val (218/72) &
  \ding{53} &
  \ding{53} &
  \checkmark &
  \ding{53} &
  \checkmark &
  \ding{53} &
  \ding{53} \\ \cline{2-9} 
 &
  Ts (73) &
  \ding{53} &
  \ding{53} &
  \ding{53} &
  \checkmark &
  \checkmark &
  \ding{53} &
  \ding{53} \\ \hline
\multirow{2}{*}{\textbf{Real US layouts}} &
  Tr/Val (218/72) &
  \checkmark &
  \ding{53} &
  \ding{53} &
  \ding{53} &
  \ding{53} &
  \ding{53} &
  \ding{53} \\ \cline{2-9} 
 &
  Ts (73) &
  \ding{53} &
  \checkmark &
  \ding{53} &
  \ding{53} &
  \ding{53} &
  \ding{53} &
  \ding{53} \\ \hline
\multirow{2}{*}{\textbf{2D CT spleen segmentations}} &
  Tr/Val (89/30) &
  \ding{53} &
  \ding{53} &
  \checkmark &
  \ding{53} &
  \checkmark &
  \checkmark &
  \ding{53} \\ \cline{2-9} 
 &
  Ts (30) &
  \ding{53} &
  \ding{53} &
  \ding{53} &
  \checkmark &
  \ding{53} &
  \ding{53} &
  \checkmark \\ \hline
\multirow{2}{*}{\textbf{Synthetic US images}} &
  Tr/Val (89/30) &
  \ding{53} &
  \checkmark &
  \checkmark &
  \ding{53} &
  \ding{53} &
  \ding{53} &
  \ding{53} \\ \cline{2-9} 
 &
  Ts (30) &
  \ding{53} &
  \checkmark &
  \ding{53} &
  \checkmark &
  \ding{53} &
  \ding{53} &
  \ding{53} \\ \hline
\multirow{2}{*}{\textbf{Synthetic US layouts}} &
  Tr/Val (89/30) &
  \ding{53} &
  \checkmark &
  \ding{53} &
  \ding{53} &
  \ding{53} &
  \ding{53} &
  \ding{53} \\ \cline{2-9} 
 &
  Ts (30) &
  \ding{53} &
  \checkmark &
  \ding{53} &
  \ding{53} &
  \ding{53} &
  \ding{53} &
  \ding{53} \\ \hline
\end{tabular}
\label{Table S1}
\end{table*}

\noindent\textbf{U-Net Training and Inference:} The segmentation U-Net was trained using the training and validation sets of both the synthetic and real US images, along with their corresponding ground truth segmentations. The learning rate was set to $10^{-3}$ with a minibatch size of 8. The model was trained for 800 epochs using the Adam optimiser. All hyperparameter values were set based on those found by optimisation as described in \cite{yuan2022deep}. During inference, the trained model predicted spleen segmentations from the test sets of both the synthetic and real US images. \\

\noindent\textbf{RVAE Training and Inference:} We first trained the RVAE to reconstruct the spleen segmentations. This pre-training process used all real US segmentations and the training/validation sets of the 2D CT-derived spleen segmentations. Note that only segmentations derived from CT volumes were used at inference (not the real US segmentations), so there was no data leakage across the training and inference. This pre-training process excluded the regression fully connected layers, and lasted for 200 epochs. Then we omitted the real US segmentations from the training set and included the regression fully connected layers for spleen volume estimation. The model was trained for a further 800 epochs. The weight of the KL-divergence was set to 0.2, and the weight for the MSE for spleen volume regression was set to 0.2. We employed the Adam optimiser with a learning rate of $10^{-3}$. We set the hyperparameter values using those found through optimisation in Section \ref{Section 4.2.1}. 

\section{Results}
\label{Section 5}
\subsection{Evaluation Metrics}
\label{Section 5.1}
To evaluate the performance of the segmentation network, we adopted the Dice score. We also computed the Hausdorff distance (HD) between the predicted segmentation contour and the ground truth segmentation contour. 

To evaluate the performance of the proposed models for spleen volume estimation, we calculated the mean relative volume accuracy (MRVA), as defined in Equation \ref{Equation 5.1}. 

\begin{equation}
    MRVA=\frac{1}{n}\sum_{i=1}^{n}\left ( 1-\frac{\left | \hat{vol}_{i}-vol_{i} \right |}{vol_{i}} \right )\times 100\%
    \label{Equation 5.1}
\end{equation}

\noindent Additionally, Pearson’s correlation coefficient (R) was calculated between the ground truth volume set $V = \{vol_1, \dots, vol_n\}$ and the estimated volume set $\hat{V} = \{\hat{vol}_1, \dots, \hat{vol}_n\}$..  

We also propose a mean confidence interval accuracy (MCIA) metric to evaluate how well the confidence limits capture the uncertainty inherent in the volume estimates. MCIA is defined as the mean of the percentage of predictions for which the ground truth volume value is within the estimated 95\% confidence limits. 

In addition, to assess the capability of our model to correctly distinguish splenomegaly cases from normal spleens, we use the classification sensitivity (SEN), specificity (SPE) and accuracy (ACC). 

\subsection{Experimental Results}
\label{Section 5.2}
\subsubsection{VAE-based Volume Estimation from 2D CT Segmentations}
\label{Section 5.2.1}
\noindent\textbf{Volume Estimation:}
For the volume estimation experiments, the results of training with single-view data (coronal slice only) and dual-view data (coronal and transverse slices) are shown in Table \ref{Table 5.1}. \textit{Human Experts} in Table \ref{Table 5.1} refers to the clinical approach of linear regression of spleen volume based on manual measurements from cross-sectional views, using either length only (for the single-view data) or the 3 measurements (for the dual-view data). 

\begin{table*}[!htb]
\caption{Comparison of results between NN, LR, RVAE, VGG-16, DenseNet-121, ResNet-18, and Pix2Vox++ trained on coronal single-view data and dual view (coronal and transverse) data. \textit{Human Experts} refers to spleen volume estimated using manual linear regression. Mean relative volume accuracy (MRVA) and standard deviation (STD). R: Pearson’s correlation coefficient. SEN: sensitivity. SPE: specificity. ACC: accuracy. Best results are in bold.}
\centering
\begin{tabular}{cccccc}
\hline
                      & \multicolumn{2}{c}{\textbf{Volume Estimation}}   & \multicolumn{3}{c}{\textbf{Splenomegaly}}          \\ \hline
                      & \textbf{MRVA$\pm$STD}               & \textbf{R} & \textbf{SEN}     & \textbf{SPE}     & \textbf{ACC} \\ \hline
\multicolumn{6}{c}{\textbf{Single view}}                                                                                      \\ \hline
\textbf{NN}           & 73.58\%$\pm$24.61\% & 0.8046     & 71.43\%          & 89.13\%          & 85.00\%      \\ \hline
\textbf{PLR}          & 53.76\%$\pm$65.60\% & 0.4742     & 60.71\%          & 83.70\%          & 78.33\%      \\ \hline
\textbf{RVAE}    & \textbf{86.62\%}$\pm$11.37\% & \textbf{0.9406 }    & \textbf{85.71\%} & 94.57\%          & \textbf{92.50\% }     \\ \hline
\textbf{VGG-16}       & 82.97\%$\pm$15.54\% & 0.9196     & \textbf{85.71\%} & 90.22\%          & 89.17\%      \\ \hline
\textbf{DenseNet-121} & 81.57\%$\pm$19.27\% & 0.9007     & \textbf{85.71\%} & 94.57\%          & \textbf{92.50\%}      \\ \hline
\textbf{ResNet-18}    & 83.10\%$\pm$17.74\% & 0.9289     & \textbf{85.71\%} & 97.83\%          & 95\%         \\ \hline
\textbf{Pix2Vox++/F}  & 82.51\%$\pm$12.33\% & 0.9244     & 67.86\%          & 97.83\%          & 90.83\%      \\ \hline
\textbf{Pix2Vox++/A}  & 82.83\%$\pm$\textbf{10.74\%} & 0.9258     & 67.86\%          & \textbf{98.91\%} & 91.67\%      \\ \hline
\textbf{Human Experts}           & 68.54\%$\pm$23.62\% & 0.8423     & -                & -                & -            \\ \hline
\multicolumn{6}{c}{\textbf{Dual views}}                                                                                       \\ \hline
\textbf{NN}           & 85.82\%$\pm$12.60\% & 0.9317     & 92.86\%          & \textbf{100\% }           & \textbf{98.33\% }     \\ \hline
\textbf{PLR}          & 84.40\%$\pm$15.70\% & 0.8773     & 71.43\%          & 96.74\%          & 90.83\%      \\ \hline
\textbf{RVAE}    & \textbf{92.58\%}$\pm$\textbf{6.07\%}  & \textbf{0.9766}     & 92.86\%          & 98.91\%          & 97.50\%      \\ \hline
\textbf{VGG-16}       & 89.96\%$\pm$8.50\%  & 0.9643     & 89.29\%          & 94.57\%          & 93.33\%      \\ \hline
\textbf{DenseNet-121} & 89.00\%$\pm$11.95\% & 0.9684     & \textbf{96.43\% }         & 97.87\%          & 97.50\%      \\ \hline
\textbf{ResNet-18}    & 91.37\%$\pm$6.31\%  & 0.9657     & 92.86\%          & 98.91\%          & 97.50\%      \\ \hline
\textbf{Pix2Vox++/F}  & 85.28\%$\pm$9.52\%  & 0.9152     & 67.86\%          & 96.74\%          & 90.00\%      \\ \hline
\textbf{Pix2Vox++/A}  & 86.27\%$\pm$10.30\% & 0.9293     & 82.14\%          & 93.48\%          & 90.83\%      \\ \hline
\textbf{Human Experts}           & 81.60\%$\pm$14.50\% & 0.9643     & -                & -                & -            \\ \hline
\end{tabular}
\label{Table 5.1}
\end{table*}

It can be seen from Table \ref{Table 5.1} that our RVAE method (MRVA 86.62\%/92.58\% for single/dual view experiments) outperformed the current clinical state-of-the-art benchmarks (MRVA 68.54\% and 81.60\% for single/dual view experiments). When compared to the comparative 2D-to-3D reconstruction-based framework Pix2Vox++ (best MRVA 82.83\% and 86.27\% for single/dual-view experiments), our RVAE method demonstrated superior volume estimation accuracy. The regression-based models (VGG-16, DenseNet-121 and ResNet-18) performed better than Pix2Vox++, but they yielded less accurate results in comparison to the RVAE method. Specifically, VGG-16 had MRVAs of 82.97\% and 89.96\%, DenseNet-121 had 81.57\% and 89.00\%, and ResNet-18 had 83.10\% and 91.37\% for single- and dual-view settings, respectively. Using paired t-tests at 0.05 significance, RVAE showed significant improvement over VGG-16 ($p=0.0065$ single view and $p=0.0053$ dual views), DenseNet-121 ($p=0.0010$ single view and $p=0.0017$ dual view) and ResNet-18 for the single view case ($p=0.0174$). Moreover, the RVAE method exhibited superior MRVA over the NN and PLR methods.

A significant observation is that results from all of our VAE-based methods as well as the regression-based models were improved when making use of dual-view data. Conversely, Pix2Vox++ did not show noticeable improvement when leveraging dual-view data compared to the other methods. There was a substantial improvement in splenomegaly prediction for all our proposed models when making use of dual-view data.

\begin{figure}[!htb] 
\centering    
\includegraphics[width=0.498\textwidth]{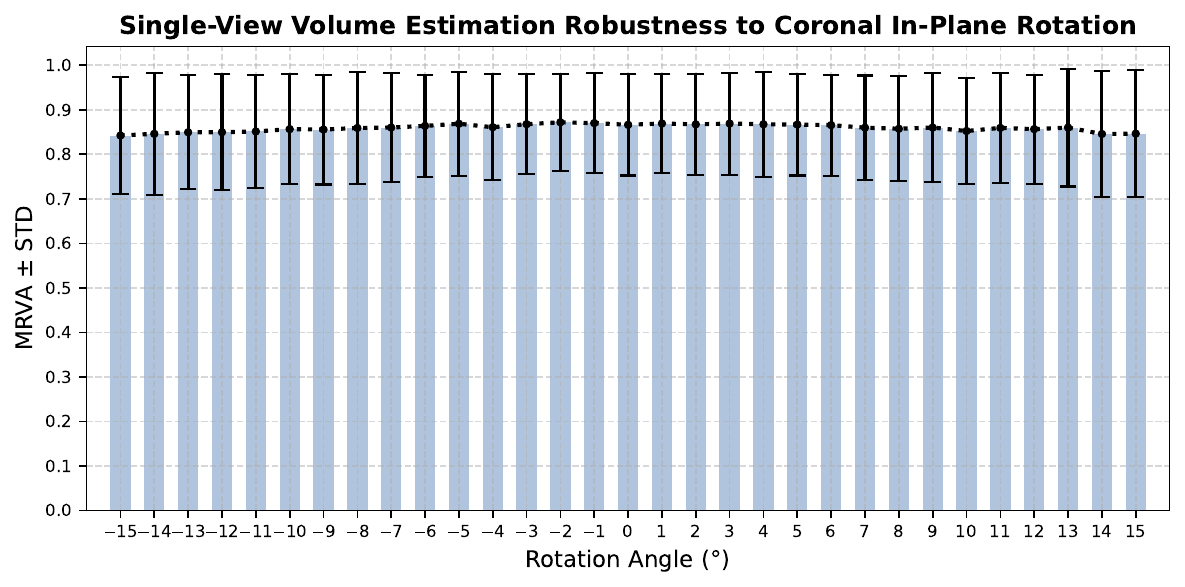}
\caption{Robustness of single-view volume estimation to coronal in-plane rotational errors.}
\label{Figure 5.1}
\end{figure}

\begin{figure}[!htb] 
\centering    
\includegraphics[width=0.498\textwidth]{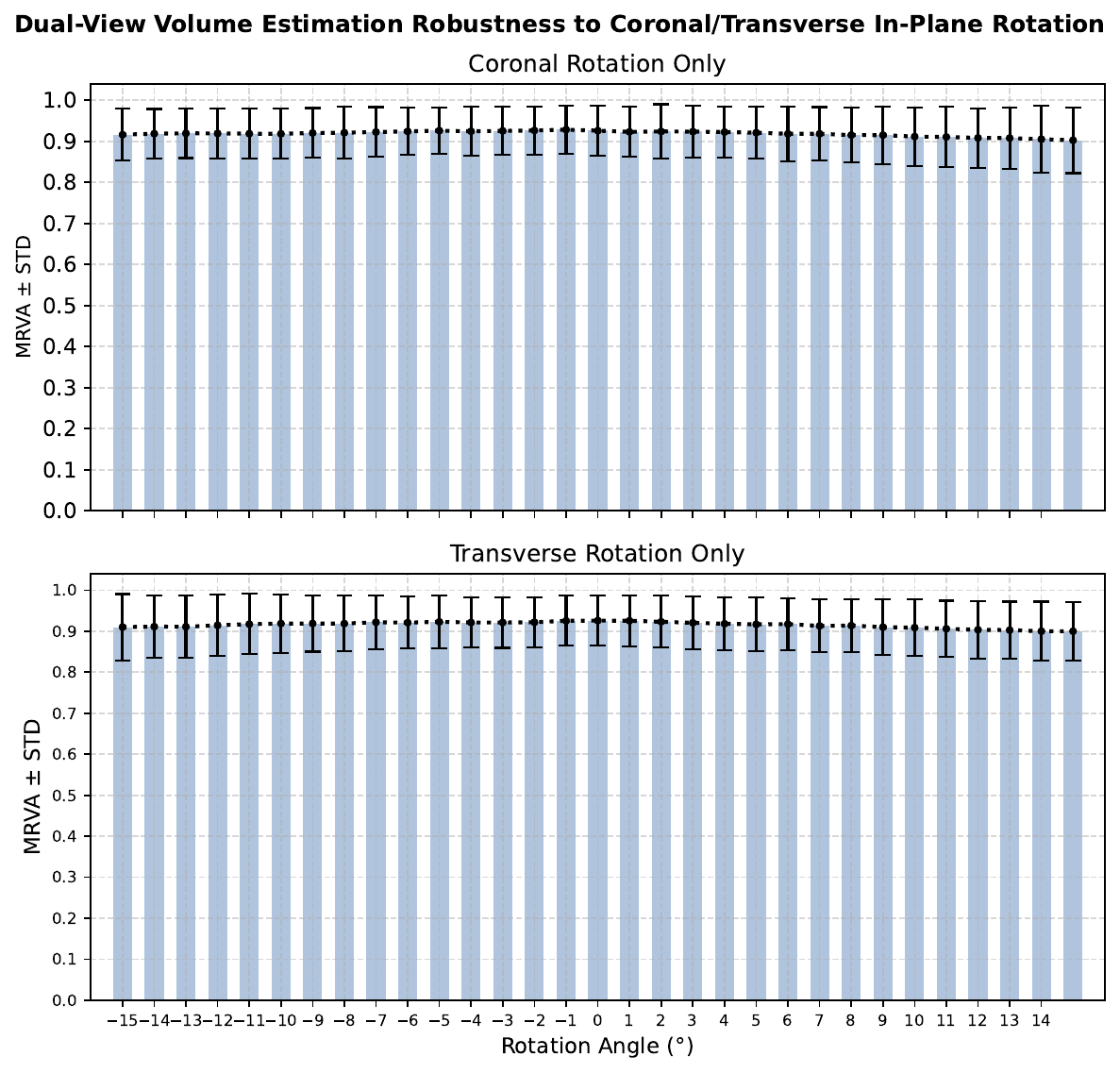}
\caption{Robustness of dual-view volume estimation to coronal and transverse in-plane rotational errors.}
\label{Figure 5.2}
\end{figure}

In addition, we conducted evaluation of our models' robustness against coronal and transverse in-plane rotational errors, which could be introduced by US probe angle changes in real clinical scenarios. The results are presented in Fig. \ref{Figure 5.1} and Fig. \ref{Figure 5.2}. It can be seen that although probe rotation can lead to a slightly increased standard deviation in volume estimation, the MRVA remains relatively stable for both single- and dual-view settings.

The results for RVAE-CI are shown in Table \ref{Table 5.2}. The model showed a similar MRVA compared to the normal RVAE model. The model trained with single-view data achieved a MCIA of 83.33\%, while the model trained with dual-view data achieved a MCIA of 84.17\%.  \\

\begin{table*}[!htb]
\caption{The results for model RVAE-CI. MRVA: mean relative volume accuracy. STD: standard deviation; R: Pearson’s correlation coefficient; MCIA: mean confidence interval accuracy; SEN: sensitivity; SPE: specificity; ACC: accuracy. }
\centering
\begin{tabular}{cccccc}
\hline
\textbf{}                           & \multicolumn{2}{c}{\textbf{Volume Estimation}} & \multicolumn{3}{c}{\textbf{Splenomegaly}}  \\ \hline
\textbf{MRVA$\pm$STD}               & \textbf{R}           & \textbf{MCIA}           & \textbf{SEN} & \textbf{SPE} & \textbf{ACC} \\ \hline
\multicolumn{6}{c}{\textbf{Single view}}                                                                                          \\ \hline
86.75\%$\pm$10.35\% & 0.931                & 83.33\%                 & 78.57\%      & 96.74\%      & 92.50\%      \\ \hline
\multicolumn{6}{c}{\textbf{Dual views}}                                                                                           \\ \hline
92.14\%$\pm$6.81\%  & 0.989                & 84.17\%                 & 85.71\%      & 98.91\%      & 95.83\%      \\ \hline
\end{tabular}
\label{Table 5.2}
\end{table*}

\begin{figure}[!htb] 
\centering    
\includegraphics[width=0.498\textwidth]{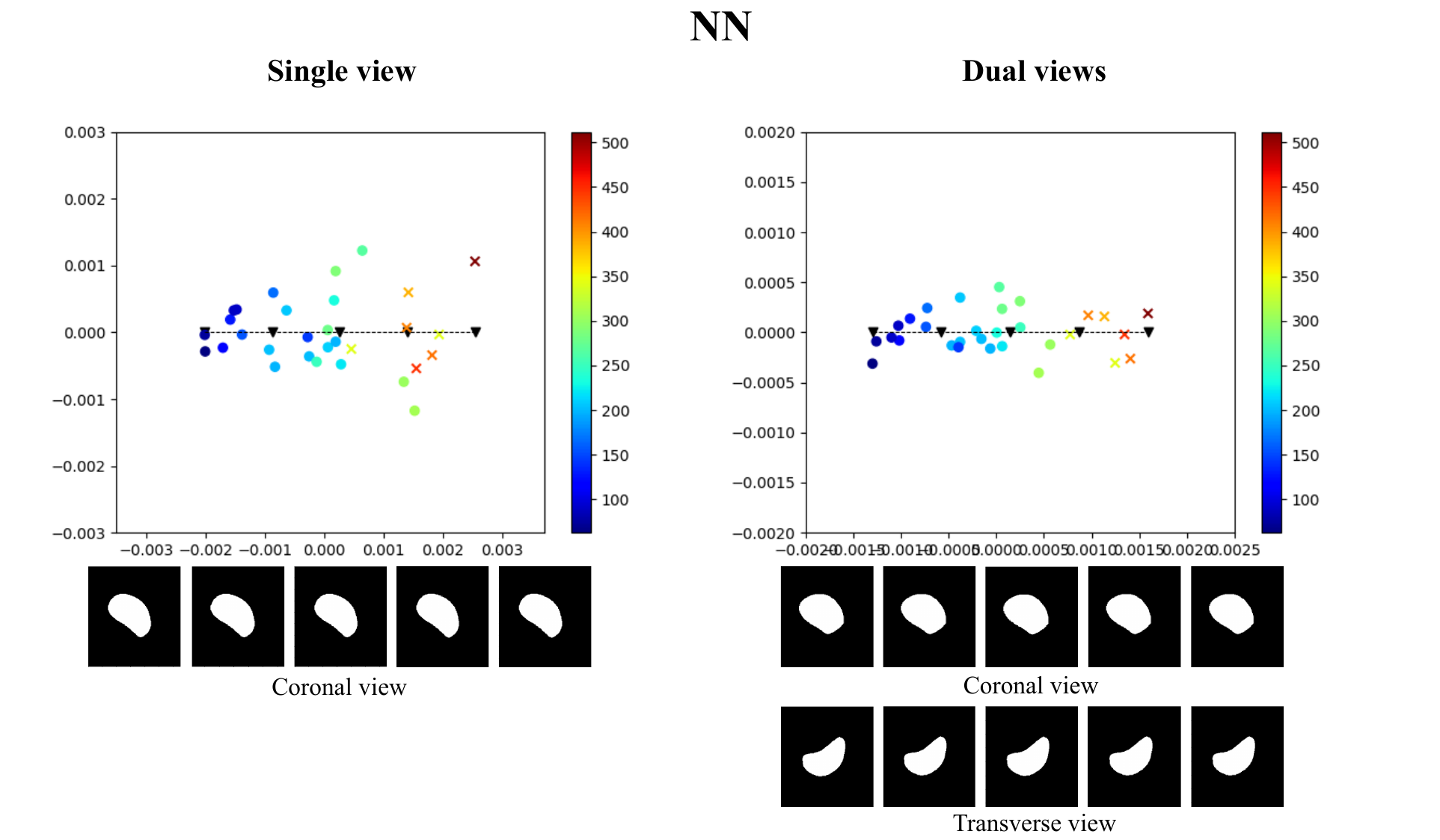}
\caption{Illustration of the test latent spaces of NN trained with single-view data or dual-view data. We sampled 5 points along the line of the principal axis, and decoded the samples to spleen segmentations, as depicted below the 2D latent maps. The colours indicate estimated spleen volume according to the colour scales shown. The dots are normal spleen cases, and crosses are splenomegaly cases (i.e., with ground truth volume value greater than 314.5mL).}
\label{Figure 5.3}
\end{figure}

\begin{figure}[!htb] 
\centering    
\includegraphics[width=0.498\textwidth]{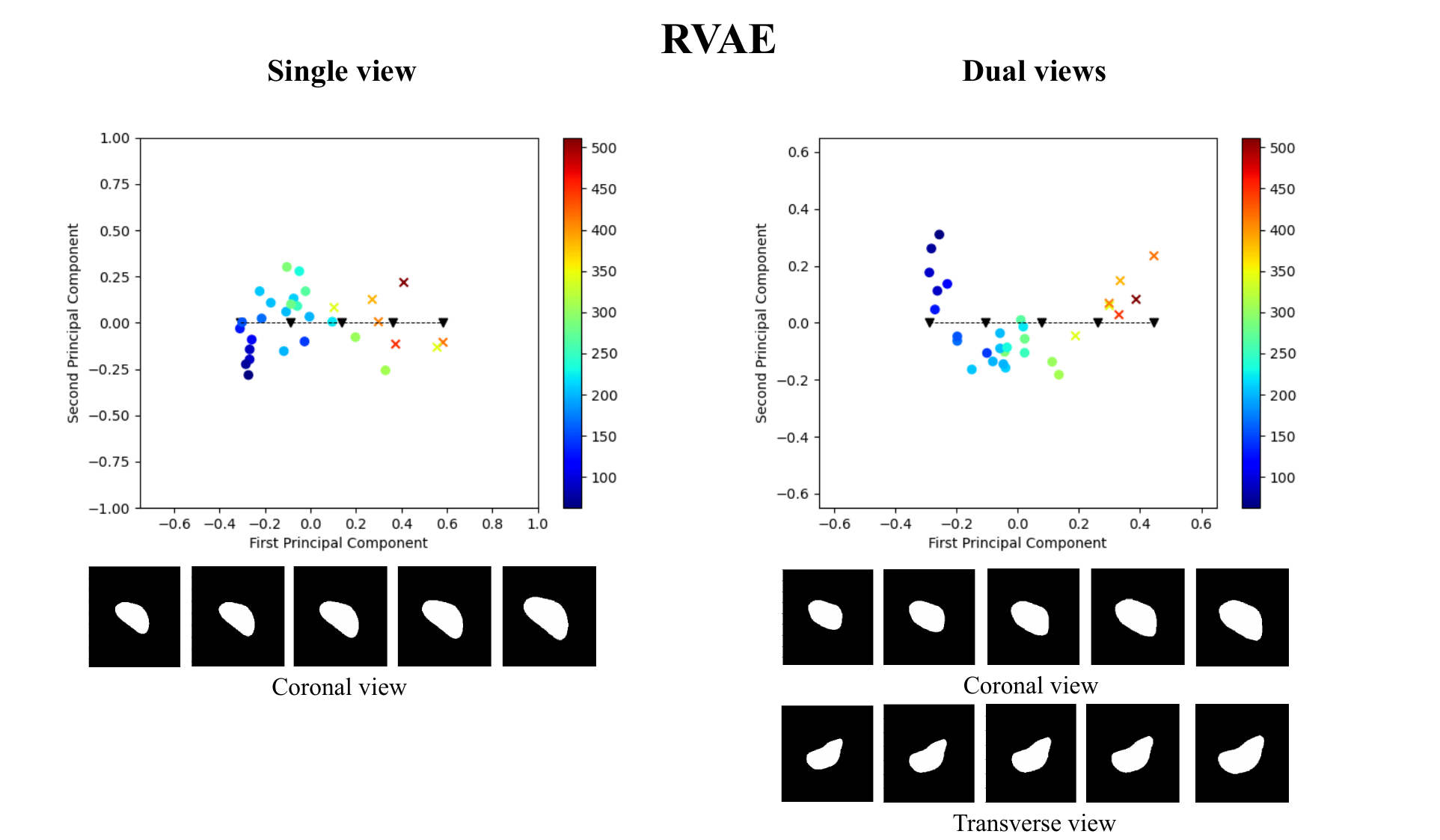}
\caption{Illustration of the test latent spaces of RVAE trained with single-view data or dual-view data. We sampled 5 points along the line of the principal axis, and decoded the samples to spleen segmentations, as depicted below the 2D latent maps. The colours indicate estimated spleen volume according to the colour scales shown. The dots are normal spleen cases, and crosses are splenomegaly cases (i.e., with ground truth volume value greater than 314.5mL).}
\label{Figure 5.4}
\end{figure}

\noindent\textbf{Interpretability:} To visualise the latent space of all proposed methods and to demonstrate the interpretability benefits of a VAE-based approach, we performed PCA on the latent representation $\mu$ and reduced its dimension from 128 to 2 for the 30 test cases, similar to \cite{puyol2020interpretable, clough2019global}. For each approach, we selected one of the four trained models in the 4-fold cross-validation and visualised the latent space under both single- and dual-view training settings. This resulted in two 2D latent maps for each proposed method. In order to sample from these latent maps and visualise the reconstructions in the original image space, we performed PCA on the latent 2D points to find the principal axis and sampled 5 points evenly along this axis. These 5 samples from the latent space were then decoded into spleen segmentations. The resulting 2D latent spaces are visualised in Fig. \ref{Figure 5.3} for NN (and PLR - note that NN and PLR share the same trained VAE, denoted simply as NN in Fig. \ref{Figure 5.3}) and Fig. \ref{Figure 5.4} for RVAE.

It can be observed from the latent space plots of all our proposed methods that normal cases and splenomegaly cases are clustered in different areas, with a clear boundary between them. There is a notable progression of the test cases in the latent map, transitioning from `low-volume' to `high volume' spleen size. When examining the spleen segmentations reconstructed from the latent space samples, the segmentations from the NN (PLR) model do not exhibit a noticeable change in spleen size. In contrast, the reconstructed spleen segmentations from the RVAE model demonstrate such progressions from smaller spleen size to larger spleen size. We further calculated Pearson's correlation coefficient R between the first principal component coordinate and corresponding spleen volume. For RVAE, the R averaged over four trained models were 0.91 and 0.96 for single- and dual-view settings, respectively. 

\subsubsection{Synthetic US Generation}
\label{Section 5.2.2}
\begin{figure}[!htb] 
\centering    
\includegraphics[width=0.48\textwidth]{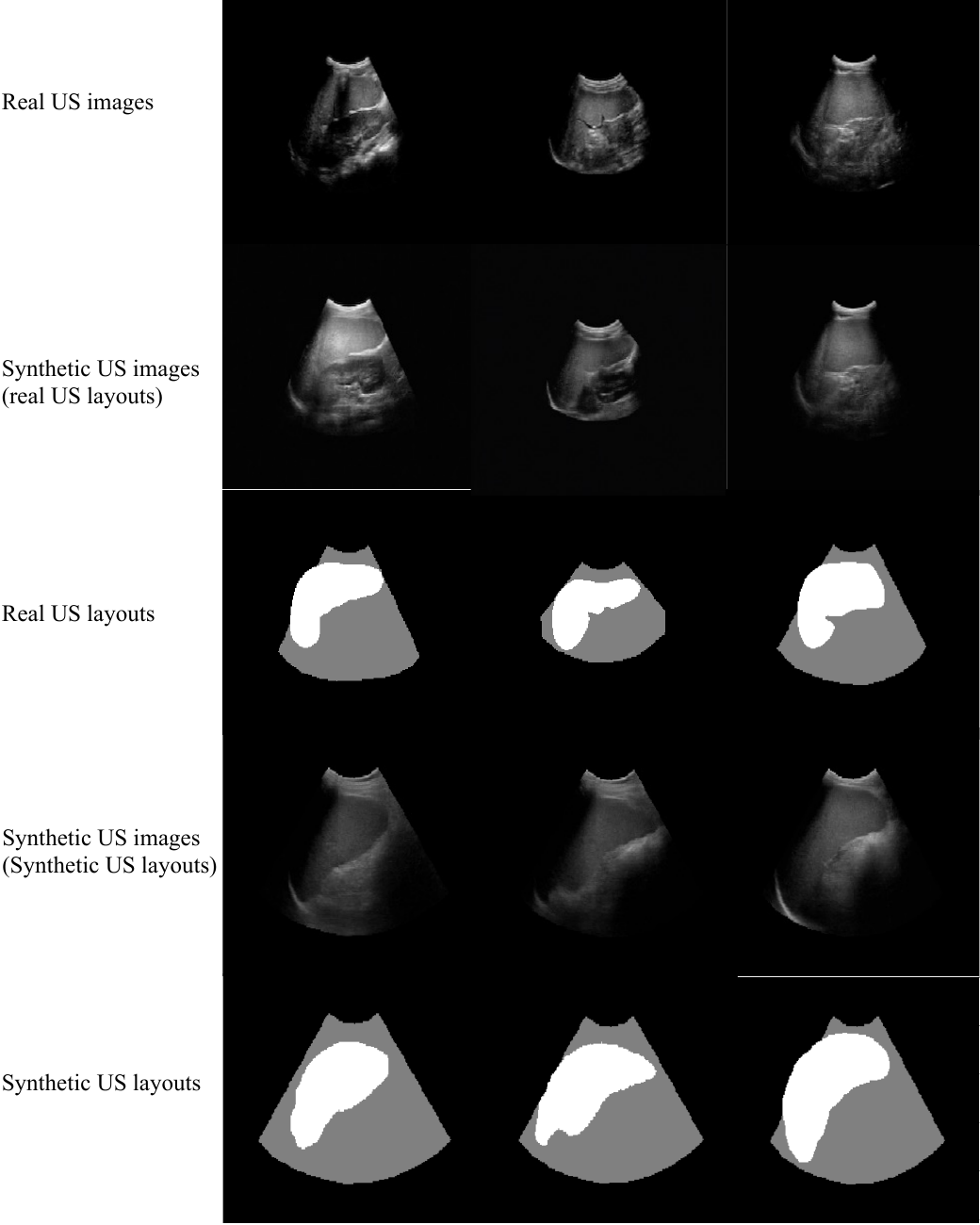}
\caption{Examples of real US images, synthetic US images generated from real US layouts, and the corresponding real US layouts. Below are shown synthetic US images and the corresponding synthetic US layouts, which were manually created based on 2D CT spleen segmentations.}
\label{Figure 5.5}
\end{figure}

We performed inference using the USDM and the US layouts produced from the CT-derived segmentations and also from real US images. Fig. \ref{Figure 5.5} illustrates the original US images alongside the corresponding synthetic US images and also the synthetic US layouts and corresponding synthetic US images. It can be seen from Fig. \ref{Figure 5.5} that the synthetic US images generated from real US layouts have strong similarity to the real US images. The USDM can simulate the tissues and structures surrounding the spleen as well as the artefacts. The synthetic US images generated from manually created US layouts also show high fidelity and realism. 

To quantitatively evaluate the similarity between the real US images and the synthetic US images generated from real US layouts, we computed the average structural similarity index (SSIM) between corresponding synthetic/real images, which was 0.7982, and the peak signal-to-noise ratio (PSNR), which was 20.98~dB. We also assessed the distributional similarity in deep feature space using the Kernel Inception Distance (KID), which was calculated as $0.0685 \pm 0.0031$. Additionally, to demonstrate its applicability in clinical tasks, we evaluated the performance of a downstream spleen segmentation task. As reported in Table~\ref{Table 5.3}, the Dice coefficient was 0.9285 for real US images and 0.9624 for synthetic US images generated from real layouts.

To further evaluate the realism of the synthetic images qualitatively, we also conducted two visual assessment experiments. In Experiment 1, 50 US images, consisting of 25 real US images and 25 synthetic US images generated from real US layouts were randomly presented to an experienced radiologist from Guy's and St Thomas' NHS Foundation Trust, who was asked to identify whether each image was synthetic or real. 13 out of 25 synthetic US images were identified as real by the radiologist, while 6 out of 25 real US images were identified as synthetic. In Experiment 2, all 149 synthetic US images generated from synthetic layouts were presented to the same radiologist who was asked to decide if they were of acceptable quality as a clinical spleen US image. Only 7 out of 149 were marked as unacceptable.

These results confirm that the synthetic ultrasound images exhibit high structural fidelity and are suitable for both visual interpretation and clinical downstream tasks.

\subsubsection{U-Net Segmentation}
\label{Section 5.2.3}
After training the U-Net with the real and synthetic US images, we tested its performance on the test sets of real and synthetic US images. The results are presented in Table \ref{Table 5.3}. 

\begin{table}[!htb]
\caption{Dice scores and HD for U-Net segmentation results on the test set of real US images, as well as results specific to synthetic US images generated using real US layouts and synthetic US layouts, respectively.}
\centering
\begin{tabular}{ccc}
\hline
                                                                                              & \textbf{Dice} & \textbf{HD} \\ \hline
\textbf{Real US images}                                                                       & 0.9285        & 11.11mm     \\ \hline
\textbf{\begin{tabular}[c]{@{}c@{}}Synthetic US images\\ (real US layouts)\end{tabular}}      & 0.9624        & 6.19mm      \\ \hline
\textbf{\begin{tabular}[c]{@{}c@{}}Synthetic US images\\ (Synthetic US layouts)\end{tabular}} & 0.9746        & 4.09mm      \\ \hline
\end{tabular}
\label{Table 5.3}
\end{table}

\begin{table*}[!htb]
\caption{MRVA for volume estimation from single-view synthetic US images using the full DeepSPV pipeline. For comparison, we also include results from our best-performing baseline model (ResNet-18, denoted as 'Baseline' below) and manual estimations by human experts.}
\centering
\begin{tabular}{ccccccc}
\hline
\multicolumn{1}{l}{} & \multicolumn{3}{c}{\textbf{Volume Estimation}} & \multicolumn{3}{c}{\textbf{Splenomegaly}} \\ \hline
\textbf{}              & \textbf{MRVA$\pm$STD}               & \textbf{R} & \textbf{MCIA} & \textbf{SEN} & \textbf{SPE} & \textbf{ACC} \\ \hline
\textbf{Baseline}      & 79.29\%$\pm$24.25\%        & 0.8918         & --            & 71.43\%           & 95.65\%           & 90.00\%           \\ \hline
\textbf{DeepSPV}       & 83.00\%$\pm$12.57\% & 0.9105     & -             & 71.43\%      & 100.00\%     & 93.33\%      \\ \hline
\textbf{DeepSPV+CI}    & 83.12\%$\pm$16.19\% & 0.9120     & 80.00\%       & 85.71\%      & 91.30\%      & 90.00\%      \\ \hline
\textbf{Human Experts} & 68.54\%$\pm$23.62\% & 0.8423     & -             & -            & -            & -            \\ \hline
\end{tabular}
\label{Table 5.4}
\end{table*}

From Table \ref{Table 5.3}, Dice scores for segmentations obtained from synthetic US images generated using real US layouts (0.9624) and synthetic US layouts (0.9746) were slightly higher compared to those obtained from real US images (0.9285). Additionally, the Hausdorff distances (HD) also indicated better performance for synthetic US images, with values of $6.19~mm$ and $4.09~mm$ for images generated from real and synthetic US layouts, respectively, compared to $11.11~mm$ for real US images.

\subsubsection{DeepSPV Full Pipeline Evaluation}
\label{Section 5.2.4}
Table \ref{Table 5.4} shows the results of the DeepSPV pipeline for single-view synthetic US images. Recall that these synthetic US images were generated from the test set of real 2D CT spleen segmentations. Therefore, we can consider these results as a fully independent assessment of the performance of the complete pipeline, comprising segmentation and volume estimation. Both DeepSPV and DeepSPV+CI perform very well (MRVAs of 83.0\% and 83.12\% respectively). As a comparison, we retrained the best-performing ResNet-18 baseline from Table~\ref{Table 5.1} with the same setup. The model achieved a MRVA of 79.29\%, which is slightly lower than the performance of our proposed method. In addition, we also present the human expert’s volume estimation using the linear regression approach \citep{bezerra2005determination}) from a single view, which is a MRVA of 68.54\%. The full pipeline volume estimation surpasses the performance of human experts by a large margin. Furthermore, it is worth noting that, by comparing the volume estimation results using ground truth segmentations (Table \ref{Table 5.1}) to those using predicted segmentations from the U-Net (Table \ref{Table 5.4}), we can see that the segmentation errors led to a modest decrease in MRVA from 86.62\% to 83.00\%.

\section{Discussion}
\label{Section 6}
In this work, we have proposed a novel framework that utilises the power of convolutional VAEs to estimate spleen volume directly from deep learning-based 2D cross-sectional segmentations. We also proposed a conditional denoising diffusion model to generate a dataset of highly realistic synthetic US images with ground truth spleen volumes. Using this dataset, we successfully evaluated the full DeepSPV volume estimation pipeline, which includes a U-Net based automated segmentation followed by a RVAE for volume estimation. The pipeline's performance surpassed the performance of a human expert using the current clinical standard approach as well as a number of comparative approaches. 

We proposed three methods for obtaining spleen volume from the latent space distribution of the VAE. The NN method estimates the volume of the spleen by finding the nearest neighbour in the latent space and uses the corresponding volume as the estimation. The PLR method is based on a linear regression between embeddings in the latent space and spleen volume. Finally, the non-linear relationship between the latent distribution embeddings and spleen volume is learned by RVAE. As shown in Table \ref{Table 5.1}, the accuracy of spleen volume estimation of the RVAE model was superior compared to NN and PLR. In addition, we evaluated the robustness of the RVAE methods against potential in-plane rotational errors introduced by incorrect probe positioning during acquisition, under both single- and dual-settings. The RVAE methods maintained stable performance with only a slight drop in MRVA, demonstrating its robustness to moderate probe misalignment.

In our DeepSPV pipeline, we chose to use VAE-based approaches to estimate volume. It is also possible to train a deep learning model to directly estimate volume without using a decoder as employed by the VAE, as we have shown in our comparative experiments on VGG-16, DenseNet-121, and ResNet-18. However, such strong supervision would likely result in a more biased representation than that learnt by the VAE, which (in its basic form) learns the representation in an unsupervised manner. It is notable that even without supervision, NN/PLR were able to capture certain clustering patterns based on spleen size in the latent space. Incorporating volume as supervision further refined the learned representation. This is also reflected in the interpretability results, where both NN/PLR and RVAE exhibited a tendency from small to large spleens in the latent space. However, the reconstructed spleen segmentations from sampled points in the RVAE model’s latent space demonstrated a more noticeable change in shape. This finding suggests that unsupervised representation learning (or less-supervised representation learning in the case of RVAE) has the potential benefit of learning a representation that could be useful for other tasks. For example, it might be possible to use the framework we have developed to predict the need for clinical intervention and/or the likelihood of further spleen enlargement. In addition, as we have shown, the decoder of the VAE allows the reconstruction of virtual spleens from different parts of the latent space distribution, which could be used for gaining new insights into the relationship between spleen morphology and patient prognosis. In terms of spleen volume estimation accuracy, our proposed RVAE approach also outperformed the more strongly supervised networks. Notably, all models were trained with the same setup, including the same maximum number of training epochs. The number of parameters in the RVAE encoder is similar to ResNet-18 (the best-performing baseline), smaller than VGG-16, and only larger than DenseNet-121.

The use of deep learning-based tools to improve existing clinical workflows has attracted increasing attention and has been widely investigated. However, the lack of interpretability in deep learning models has been one of the major concerns when applied to real clinical scenarios. Despite their accuracy, most deep learning models remain black boxes, making it difficult for clinicians to understand how the outcome is obtained. This lack of interpretability can lead to a lack of trust in clinical decision-making. To alleviate this concern, we have provided visualisations of the model’s latent space so that clinicians can see how the representation learnt by the model is closely related to volume. We observed that there is a clear boundary between splenomegaly cases and normal cases and a clear progression from small cases to large cases in the latent space. We believe these should improve clinician trust in the models. In addition, to provide more information to support clinicians, we developed a technique to quantify confidence intervals for the volume estimates, rather than just providing a single scalar value. In this approach, the regression layers made predictions from multiple samples in the latent distribution. This process is analogous to the uncertainty that arises from obtaining a 2D slice from a 3D spleen volume, where the resulting 2D segmentation can vary due to motion or slight differences in orientation. The confidence intervals could also serve to increase clinical trust in the volume estimates and could even be used for automated detection of splenomegaly cases with a level of uncertainty.

As mentioned in Section \ref{Section 1.2}, it would be possible to estimate spleen volume from 3D imaging such as CT. However, 3D imaging is usually not deployed to measure spleen volume as it is more expensive and less accessible. Practically, estimating the spleen volume from 2D imaging is more clinically useful as it can be incorporated into routine clinical workflows. In this setting, reconstructing the 3D segmentation from 2D segmentation slices and then calculating the spleen volume is a natural choice. Such a method was proposed and demonstrated in cardiac imaging by \cite{stojanovski2022efficient}. However, our results showed that the Pix2Vox++ 2D-to-3D reconstruction methods did not perform as well as our direct volume estimation approach. Besides, the reconstruction accuracy/volume estimation accuracy of 2D-to-3D reconstruction techniques depends on having sufficient geometrical information. Considering the sparseness of slice sampling in clinically acquired US of the spleen, the reconstructed volume will not be reliable and therefore would be of questionable clinical value. Another possibility would be to use multiple 3D US volumes or mechanical sweeps of 2D US images acquired under breath-hold. Future work could focus on addressing the associated limitations and challenges and extending our framework to enable splenic volume estimation from such data.

In this study, we tackled the challenge of the lack of paired 2D US and ground truth volumes by utilising a diffusion model to generate a synthetic US image dataset from 2D CT segmentations. Our USDM demonstrated a high level of fidelity and realism in synthetic spleen US image generation. Provided with real US layouts, it can accurately generate synthetic US images which closely match their real counterparts. From a qualitative inspection of Fig. \ref{Figure 5.5}, the synthetic US images generated based on real US layouts can effectively emulate real-world US image content. Quantitative evaluation supported this observation. However, although the synthetic US images generated from manually created US layouts match the appearance of real spleen US images, they do occasionally lack variation and details of the surrounding tissues when compared to the images generated from real US layouts. This influenced the segmentation results, where the Dice and HD for those generated from manually created US layouts are higher. In future work, having more realistic US layouts could mitigate this limitation, but it is possible that the domain gap between US and CT derived segmentations cannot be entirely eliminated.

While our volume estimation network has demonstrated superior performance on 2D single- and dual-view-based volume estimation when compared to the current clinical workflow, a limitation of our work is that we were only able to generate synthetic 2D coronal US images and evaluate the full pipeline under single-view conditions due to the lack of transverse 2D US images for training the USDM. The full pipeline evaluation on single-view 2D synthetic US images yielded similar accuracy to that obtained from ground-truth 2D CT segmentations (MRVA: 83.00\% vs. 86.62\%), with both surpassing human expert measurements using current standard clinical practice (68.54\%). However, this result indicates that errors from the automated segmentation process would have an impact on volume estimation accuracy (a decrease of 3.62\% in MRVA). This highlights the importance of segmentation accuracy to the final volume estimation. Further investigation with real-world paired datasets may help to better quantify the robustness and understand the accumulation of such errors. With transverse 2D US images acquired in future studies, the USDM can be fine-tuned to generate synthetic transverse images, thus enabling evaluation of the full pipeline in dual-view scenarios. Additionally, future collection of paired US and CT datasets would allow further validation of our methods on real clinical data. Finally, establishing clinically acceptable thresholds for volumetric measurement errors remains an open challenge, which would require future large-scale clinical validation studies.

It should be acknowledged that the CT data we used were obtained from adults, while the US data were acquired from paediatric SCD patients. It is important to note that the spleen grows rapidly during the first five years of life and typically stabilises around the age of 13 \citep{megremis2004spleen, prassopoulos1994ct,niederau1983sonographic}. For example, typical normal upper limits of spleen length (measured from 2D US) are 11.5cm at 12 years and 12-13cm at age 15 or older \citep{rosenberg1991normal}. Additionally, spleen size varies and is not always strictly stratified by children or adults in clinical assessments. Moreover, our US datasets were from SCD patients, some of whom manifested splenomegaly. Given these considerations, we do not expect this difference to have a significant impact on our results and conclusions. However, with more datasets acquired in the future, stricter quality control in terms of age stratification and pathology will be necessary to further validate the generalisability of our approach.

In this work, our aim was to develop new technology that could be integrated into existing clinical workflows using already-acquired data. The motivation of developing this pipeline primarily derived from three main clinical needs. Firstly, spleen volume has the potential to play a more important role in diagnosing complex disease-related splenomegaly and gauging the severity of related diseases compared to spleen length. Secondly, utilising a 3D imaging modality like CT for spleen assessment involves ionising radiation and is less accessible in low and middle income countries, such as many countries in the Global South where there is a high prevalence of SCD-related splenomegaly. Lastly, US examination is the most widely used assessment of the spleen and typically only coronal and transverse US images are acquired. Drawing from these motivations, we devised a pipeline which consists of segmentation and volume estimation networks to estimate the spleen volume from routinely acquired US images. When introducing new technology into clinical workflows, trust plays a crucial role, and we acknowledge that additional validation needs to be done before clinicians can fully rely on volume estimates made from single- or dual-view US images. We hope that the interpretability results presented in this paper will serve as an initial step towards building that trust. Nevertheless, we believe that, if deployed clinically, our method would initially be utilised alongside the current standard approach of length measurement. Over time, as confidence in the system grows, clinicians may appreciate the added value of our technique, particularly its ability to provide 3D information whilst also removing the need for manual measurement.

\section{Conclusion}
\label{Section 7}
In this work, we have proposed DeepSPV, consisting of a 2D US segmentation model and a volume estimation model, to estimate the spleen volume from 2D spleen US images. To the best of our knowledge, this is the first work employing 2D US for 3D spleen volume estimation and has surpassed human expert level and existing reconstruction-based and regression-based methods in spleen volume estimation. 

\section*{Data Availability}

The dataset of synthetic 2D US images and corresponding ground truth spleen volumes will be made publicly available on paper publication.

\section*{Acknowledgments}
We acknowledge funding from the Wellcome/EPSRC Centre for Medical Engineering at King’s College London (WT 203148/Z/16/Z), the National Institute for Health Research (NIHR) Cardiovascular MedTech Co-operative award to the Guy’s and St Thomas’ NHS Foundation Trust, and the NIHR comprehensive Biomedical Research Centre award to Guy’s \& St Thomas’ NHS Foundation Trust in partnership with King’s College London. The support provided by the China Scholarship Council for the PhD programme of Zhen Yuan at King’s College London is also acknowledged.

\bibliographystyle{model2-names.bst}\biboptions{authoryear}
\bibliography{refs}

\end{document}